\documentstyle[12pt]{article}

\makeatletter
\@addtoreset{equation}{section}
\makeatother

\let\la=\label  
   
 \def\bd{\begin{document}} \def\ed{\end{document}}
\def\ds{\documentstyle} \let\fr=\frac \let\bl=\bigl \let\br=\bigr
\let\Br=\Bigr \let\Bl=\Bigl 
\let\bm=\bibitem
\let\na=\nabla
\let\pa=\partial \let\ov=\overline 
\newcommand{\be}{\begin{equation}} 
\newcommand{\ee}{\end{equation}} 
\def\ba{\begin{array}}
\def\ea{\end{array}}
\newcommand{\ho}[1]{$\, ^{#1}$}
\newcommand{\hoch}[1]{$\, ^{#1}$}
\newcommand{\bea}{\begin{eqnarray}} 
\newcommand{\eea}{\end{eqnarray}} 
\newcommand{\ra}{\rightarrow}
\newcommand{\lra}{\longrightarrow}
\newcommand{\Lra}{\Leftrightarrow}
\newcommand{\ap}{\alpha^\prime}
\newcommand{\bp}{\tilde \beta^\prime}
\newcommand{\tr}{{\rm tr} }
\newcommand{\Tr}{{\rm Tr} } 
\newcommand{\NP}{Nucl. Phys. }
\newcommand{\tamphys}{\it Center for Theoretical Physics\\
Texas A\&M University, College Station, Texas 77843}

\newcommand{\auth}{M. J. Duff\footnote{Research supported in  part by 
NSF Grant PHY-9411543.}}

\begin{document}

\hfill{CTP-TAMU-61/96}

\hfill{hep-th/9611203}

\vspace{24pt}

\begin{center}
{ \large 
{\bf SUPERMEMBRANES\footnote{Based on lectures given at 
the T. A. S. I. Summer School, University of Colorado, Boulder, June 1996;
the Topical Meeting, Imperial College, London, July 1996 and the 26th
British Universities Summer School in Theoretical Elementary Particle
Physics, University of Swansea, September 1996.}}}

\vspace{36pt}

\auth

\vspace{10pt}

{\tamphys}

\vspace{44pt}

\underline{ABSTRACT}

We give an elementary introduction to the theory of supermembranes.
\end{center}
\bigskip
\bigskip
\bigskip
\bigskip
\bigskip
\bigskip
\bigskip
\begin{center}
{\it DEDICATED TO THE MEMORY OF ABDUS SALAM}
\end{center}
{\vfill\leftline{}\vfill}

\pagebreak
\setcounter{page}{1}

\tableofcontents
\newpage

\section{Supermembranes}
\label{super}

Membrane theory has a strange history which goes back even further than
strings. The idea that the elementary particles might correspond to
modes of a vibrating membrane was put forward originally in 1962 by 
Dirac \cite{Dirac}.  When string theory came along in the 1970s, there were 
some attempts to revive the membrane idea but without much success.
Things did not change much until 1986 when Hughes, Liu and Polchinski
\cite{Hughes} showed that it was possible to combine membranes with
supersymmetry: the {\it supermembrane} was born.

Consequently, while all the progress in string theory was going on, a small
splinter group was posing the question: Once you have given up
$0$-dimensional particles in favor of $1$-dimensional strings, why not
$2$-dimensional membranes or in general $p$-dimensional objects (inevitably
dubbed {\it $p$-branes})?  Just as a $0$-dimensional particle sweeps out a
$1$-dimensional {\it worldline} as it evolves in time, so a $1$-dimensional
string sweeps out a $2$-dimensional {\it worldsheet} and a $p$-brane sweeps
out a $d$-dimensional {\it worldvolume}, where $d=p+1$.  Of course, there
must be enough room for the $p$-brane to move about in spacetime, so $d$
must be less than or equal to the number of spacetime dimensions $D$.  In
fact, as we shall see in section (\ref{scan}) supersymmetry places further
severe restrictions both on the dimension of the extended object and the
dimension of spacetime in which it lives \cite{Achucarro}. One can
represent these as points on a graph where we plot spacetime dimension $D$
vertically and the $p$-brane dimension $d=p+1$ horizontally. This graph is
called the {\it brane scan}. See Table \ref{branescan}.  In the early
eighties Green and Schwarz \cite{Greenschwarz} had shown that spacetime
supersymmetry allows classical superstrings moving in spacetime dimensions
$3,4,6$ and $10$. (Quantum considerations rule out all but the
ten-dimensional case as being truly fundamental. Of course some of these
ten dimensions could be curled up to a very tiny size in the way suggested
by Kaluza and Klein \cite{Kaluza,Klein}. Ideally six would be compactified
in this way so as to yield the four spacetime dimensions with which we are
familiar.) It was now realized, however, that these $1$-branes in
$D=3,4,6$ and $10$ should now be viewed as but special cases of this more
general class of supersymmetric extended object.

\begin{table}
\caption{ The brane scan, where $S$, $V$ and $T$ denote scalar, vector and
antisymmetric tensor multiplets.}
$
\begin{array}{ccccccccccccccc}
~&D\uparrow&&&&&&&&&&&~\\
~&11&.&~&&S&&&T&&&&?&~\\
~&10&.&V&S/V&V&V&V&S/V&V&V&V&V&~\\
~&9&.&S&&&&S&&&&&&~\\
~&8&.&~&&&S&&&&&&&~\\
~&7&.&~&&S&&&T&&&&&~\\
~&6&.&V&S/V&V&S/V&V&V&&&&&~\\
~&5&.&S&&S&&&&&&&&~\\
~&4&.&V&S/V&S/V&V&&&&&&&~\\
~&3&.&S/V&S/V&V&&&&&&&&~\\
~&2&.&S&&&&&&&&&&~\\
~&1&.&~&~&~&~&~&~&~&~&~&~&~\\
~&0&.&.&.&.&.&.&.&.&.&.&.&.~\\
~&~&0&1&2&3&4&5&6&7&8&9&10&11&d\rightarrow
\end{array}
$
\label{branescan}
\end{table}

Curiously enough, the maximum spacetime dimension permitted is eleven, 
where
Bergshoeff, Sezgin and Townsend found their supermembrane
\cite{Bergshoeff1,Bergshoeff2} which couples to eleven-dimensional
supergravity \cite{Cremmer,Kaluza}. (The $3$-form gauge field of $D=11$
supergravity had long been suggestive of a membrane interpretation
\cite{Julia}). Moreover, it was then possible to show \cite{Howe} by
simultaneous dimensional reduction of the spacetime and worldvolume 
that the membrane looks like a string in ten dimensions. In fact, it yields
precisely the Type $IIA$ superstring. This suggested that the
eleven-dimensional theory was perhaps the more fundamental after all.  

Notwithstanding these and subsequent results, the supermembrane enterprise
was, until recently, largely ignored by the mainstream physics community.
Those who had worked on eleven-dimensional supergravity  and
then on supermembranes spent the early eighties arguing for {\it spacetime}
dimensions greater than four, and the late eighties and early nineties
arguing  for {\it worldvolume} dimensions greater than two. The latter
struggle was by far the more bitter!   

In these lectures we shall review the progress reached over the last
decade and see how it fits in with recent results in string duality,
$D$-branes \cite{Polchinski} and $M$-theory
\cite{Howe,Luduality,Hulltownsend,Townsendeleven,Wittenvarious,%
Duffliuminasian,Becker1,Schwarzpower,Aharony,BanksM}. I shall draw
on material contained in previous reviews on supermembranes
\cite{Duffsuper,Duffclassical,Duffpopesezgin,Khuristring}, duality
\cite{Duffelectric} and $M$-theory \cite{DuffM}.  The reader is also
referred to the reviews by Townsend on supermembranes \cite{Townsendrecent}
by Polchinski on $D$-branes \cite{TASI} and by Schwarz \cite{Schwarzreview}
on $M$-theory.

\subsection{Bosonic $p$-branes}
\label{bose}
Consider some extended object with $1$ time and $(d-1)$ space dimensions
moving in a spacetime with $1$ time and $(D-1)$ space dimensions.  We 
shall demand that its dynamics is governed by minimizing the worldvolume
which the object sweeps out
\begin{equation}
S = -T_d \int d^d \xi~\lbrace- det\partial_i x^{\mu}~\partial_j x^{\nu}
\eta_{\mu\nu}\rbrace^{1/2}
\end{equation}
where we have introduced worldvolume coordinates $\xi^i$ (i = 0, $\ldots$,
(d-1)) and spacetime coordinates $x^{\mu}$ ($\mu$ = 0, $\ldots$, (D-1)). 
To begin with, we assume spacetime is flat with Minkowski metric $\eta_{\mu
\nu}$ and signature $(-, +, \ldots, +)$.  The tension of the object is
given by the constant $T_d$ which renders the action $S$ dimensionless. 
This action was  first introduced by Dirac \cite{Dirac} in the case of a
membrane $(d = 3)$ and later by Nambu \cite{Nambu} and Goto \cite{Goto} in
the case of a string $(d = 2)$. 

The classical equations of motion that follow from (1.1) may equivalently be
obtained from the action
\begin{equation}
S =  T_d\int d^d \xi \biggl( -\frac{1}{2}\sqrt{-\gamma} 
\gamma^{ij} \partial_i
x^{\mu} \partial_j x^{\nu} \eta_{\mu\nu} +\frac{1}{2}(d-2) {\sqrt -\gamma}
\biggr) 
\end{equation}
where, following Howe and Tucker \cite{Howetucker} and Polyakov
\cite{Polyakov}, we have introduced  the auxiliary field
$\gamma_{ij}(\xi)$.  $\gamma$ denotes its determinant and $\gamma^{ij}$ its
inverse.  Varying with respect to $\gamma_{ij}$ yields the equation of motion
\begin{equation}
\frac{1}{2} \sqrt{-\gamma}\gamma^{ij}\gamma^{k\ell}\partial_k x^{\mu}
\partial_{\ell} x^{\nu}\eta_{\mu\nu}-\sqrt{-\gamma}\partial_k
x^{\mu}\partial_{\ell} x^{\nu}\gamma^{ik}\gamma^{j\ell}
=\frac{1}{2} (d - 2)\sqrt{-\gamma} \gamma^{ij}~~.
\end{equation}
Taking the trace, we find for $d \neq 2$, that
\begin{equation}
\gamma^{k\ell} \partial_k x^{\mu} \partial_{\ell} x^{\nu}~ 
\eta_{\mu \nu} =d  
\end{equation}
and hence that $\gamma_{ij}$ is just the induced metric on worldvolume
\begin{equation}
\gamma_{ij} = \partial_i x^{\mu} \partial_j x^{\nu} \eta_{\mu \nu} . 
\end{equation}
Varying (1.2) with respect to $x^{\mu}$ yields
\begin{equation}
\partial_i \left( \sqrt{-\gamma} \gamma^{ij} \partial_j x^{\nu} 
\eta_{\mu \nu}
\right) = 0.
\end{equation}
Thus equations (1.5) and (1.6) are together equivalent to the equation of
motion obtained by varying (1.1) with respect to $x^{\mu}$.

Note that the case $d = 2$ is special.  Here, the worldvolume
cosmological term drops out and (1.2) displays a conformal symmetry
\begin{eqnarray}
&\gamma_{ij}(\xi) \rightarrow \Omega^2 (\xi) \gamma_{ij}
(\xi)\nonumber\\
&x^{\mu}(\xi) \rightarrow x^{\mu} (\xi) 
\end{eqnarray}
where $\Omega$ is some arbitrary function of $\xi$.  In this case 
$\gamma_{ij}$ and $\partial_i x^{\mu} \partial_j x^{\nu} \eta_{\mu \nu}$
are related only up to a conformal factor. The actions (1.1) and (1.2) are,
however, equivalent for all $d$, at least classically.  As discussed in
\cite{Deser}, it is possible to construct a conformally invariant action for
all $d$, by the simple expedient of raising the usual Lagrangian to the
power $d/2$
\begin{equation}
S = -T_d \int d^d \xi \sqrt{-\gamma} \left( \frac{1}{d} \gamma^{ij}
\partial_i x^{\mu} \partial_j x^\nu \partial_j x^{\nu} \eta_{\mu \nu}
\right) ^{d/2}~~.  
\end{equation}
The equations of motion read
\begin{eqnarray}
 &&\frac{1}{2} \sqrt{-\gamma} \gamma^{ij}\left( \frac{1}{d}\gamma^{k\ell}
\partial_k x^{\mu} \partial_{\ell}
x^{\nu}\eta_{\mu\nu}\right)^{d/2}
\nonumber \\
 && ~~~~~~~~~~
=\frac{1}{2}\biggl( \frac{1}{d}\gamma^{k\ell}\partial_k
x^{\mu}\partial_{\ell}x^{\nu}\eta_{\mu\nu}\biggr)^{d/2-1}
\gamma^{im}\gamma^{jn}\partial_m x^{\rho} \partial_n x^{\sigma}
\eta_{\rho\sigma}   
\end{eqnarray}
and
\begin{equation}
\partial_i \left[ \sqrt{-\gamma} \left( \gamma^{k\ell} \partial_i x^{\mu}
\partial_j x^{\sigma} \eta_{\rho \sigma} \right) ^{d/2 -1} \gamma^{ij}
\partial_j x^{\nu} \eta_{\mu \nu} \right] = 0 
\end{equation}
which are again equivalent to the Dirac and Howe-Tucker equations of
motion.

There are two useful generalizations of the above.  The first is to go to
curved space by replacing $\eta_{\mu \nu}$ by $g_{\mu \nu} (x)$; the second
is to introduce an antisymmetric tensor field $B_{\mu \nu \ldots \rho} (x)$
of rank $d$ which couples via a Wess-Zumino term.  The action (1.2) then
becomes
\[
S = T_d \int d^d \xi \biggl[ -\frac{1}{2}\sqrt{-\gamma} \gamma^{ij}
\partial_i x^{\mu} \partial_j x^{\mu} g_{\mu\nu} (x) +\frac{1}{2}(d - 2)
\sqrt{-\gamma}
\]
\begin{equation} 
+\frac{1}{d!} \epsilon^{i_{1} i_{2} .. i_{d}}
\partial_{i_{1}} x^{\mu_1} \partial_{i_{2}} x^{\mu_2} \ldots
\partial_{i_{d}} x^{\mu_d} B_{\mu_{1} \mu_{2} .. \mu_d} (x)
\biggr]   
\end{equation}
and the equations of motion are
\begin{eqnarray}
 &&
\partial_i \left( \sqrt{-\gamma} \gamma^{ij} \partial_j x^{\nu} 
g_{\mu \nu}\right)+
g_{\mu \rho} 
\Gamma^{\rho}{}_{\kappa\lambda}\partial_ix^{\kappa}\partial_j x^{\lambda}
\gamma^{ij}
 \nonumber \\
 && ~~~~~~~~~~~ =
\frac{1}{d!}F_{\mu\nu\tau\cdots\sigma}\epsilon^{ij\cdots
k}\partial_ix^{\nu}\partial_jx^{\tau}
\ldots\partial_kx^{\sigma} 
\end{eqnarray}
and
\begin{equation}
\gamma_{ij}=\partial_ix^{\mu}\partial_jx^{\nu}g_{\mu\nu}(x)
\end{equation}
where the field-strength $F$ is given by
\begin{equation}
F=dB
\end{equation}
and hence obeys the Bianchi identity
\begin{equation}
dF = 0.
\end{equation}
The virtue of these generalizations is that they now permit a 
straightforward transition to the supermembrane.

Our experience with string theory suggests that there are two ways of
introducing supersymmetry into membrane theory.  The first is to look
for a {\it supermembrane} for which has manifest spacetime supersymmetry but
no supersymmetry on the worldvolume.  The second is to look for a
{\it spinning membrane} which has manifest worldvolume supersymmetry but no
supersymmetry in spacetime.  An early attempt at spinnning membranes by
Howe and Tucker \cite{Howetucker} encountered the problem that the
worldvolume cosmological term does not permit a supersymmetrization using
the usual rules of $d = 3$ tensor calculus without the introduction of an
Einstein-Hilbert term \cite{Inami}.  Indeed, these objections have been
elevated to the status of a {\it no-go theorem} for spinning membranes
\cite{Bergshoeff3}. An attempt to circumvent this no-go theorem was made 
by Lindstrom and Rocek \cite{Lindstrom} starting from the conformally
invariant action (1.8), but is is fair to say that the spinning membrane
approach never really caught on. Recently, there has been some success in
formalisms with both worldvolume and spacetime supersymmetry
\cite{Howesezgin,Howesezginp}. The last ten years of supermembranes,
however, has been dominated by the approach with spacetime supersymmetry 
and worldvolume kappa symmetry.  At first, progress in
supermembranes was hampered by the belief that kappa symmetry,  so
crucial to Green-Schwarz superparticles ($d=1$) \cite{Siegel} and
superstrings ($d=2$) \cite{Greenschwarz} could not be generalized to
membranes.  The breakthrough came when Hughes, Liu and Polchinski
\cite{Hughes} showed that it could. 

\subsection{Super $p$-branes}
\label{fermi}

It is ironic that although one of the motivations for the original
supermembrane paper \cite{Hughes} was precisely to find the superthreebrane
as a {\it topological defect} of a supersymmetric field theory in $D=6$; 
the discovery of the other supermembranes proceeded in the opposite
direction. Hughes et al. showed that kappa symmetry could be generalized
to $d > 2$ and proceeded to construct a threebrane displaying an explicit
$D=6, N=1$ spacetime supersymmetry and kappa invariance on the
worldvolume. It was these kappa symmetric Green-Schwarz actions, rather
then the soliton interpretation which was to dominate the early work on the
subject. First of all, Bergshoeff, Sezgin and Townsend \cite{Bergshoeff1}
found corresponding Green-Schwarz actions for other values of $d$ and $D$,
in particular the eleven-dimensional supermembrane. (We shall discuss the
$D=11$ supermembrane in section (\ref{reduce}) and show how to derive from it
the Type $IIA$ string in ten dimensions by a simultaneous dimensional
reduction of the worldvolume and the spacetime.) Their method was to
show that such Green-Schwarz super $p$-brane actions are possible whenever
there is a closed $(p+2)$-form in superspace. As described in section
(\ref{scan}), the four ``fundamental'' super $p$-branes are then given by
$p=2$ in $D=11$, $p=5$ in $D=10$, $p=3$ in $D=6$ and $p=2$ in $D=4$
\cite{Achucarro}. In fact, applying the above mentioned simultaneous reduction $k$
times, we find four sequences of $(p-k)$-branes in $(D-k)$ dimensions, which
include the well known Green-Schwarz superstrings in $D=10,6,4$ and $3$.
These four sequences, known as the octonionic (${\cal O}$), quaternionic
(${\cal H}$), complex (${\cal C}$) and real (${\cal R}$) sequences are
desribed by the points labelled $S$ on the brane scan of Table
\ref{branescan}. 

Let us introduce the coordinates $Z^M$ of a curved superspace
\begin{equation}
Z^M=(x^{\mu}, \theta^{\alpha})
\end{equation}
and the supervielbein $E_M{}^A (Z)$ where $M = \mu,\alpha$ are world
indices and $A = a,\alpha$ are tangent space indices.  We also define the
pull-back
\begin{equation}
E_i{}^A = \partial_iZ^ME_M{}^A~~.
\end{equation}
We also need the super-$d$-form $B_{A_{d}\ldots A_{1}}(Z)$. 
Then the supermembrane action is 
\begin{eqnarray}
S= T_d\int
d^d\xi\biggl\lbrack &-&\frac{1}{2}{\sqrt -\gamma}\gamma^{ij}
E_i{}^aE_j{}^b \eta_{ab}+\frac{1}{2} (d-2){\sqrt -\gamma}
\nonumber \\
&+&\frac{1}{d!}\epsilon^{i_1\ldots
i_d}E_{i_{1}}{}^{A_1}\cdots E_{i_{d}}{}^{A_d}B_{A_{d}\ldots A_{1}}
\biggr\rbrack.
\label{supermembrane}
\end{eqnarray}
As in (1.11) there is a kinetic term, a worldvolume cosmological term, and
a Wess-Zumino term.  The action (1.18) has the virtue that it reduces to
the Green-Schwarz superstring action when $d = 2$.

The target-space symmetries are superdiffeomorphisms, Lorentz invariance and
$d$-form gauge invariance.  The worldvolume symmetries are ordinary
diffeomorphisms and kappa invariance referred to earlier which is known
to be crucial for superstrings, so let us examine it in more detail.  The
transformation rules are

\begin{equation}
\delta Z^M E^a{}_M = 0, ~~~\delta Z^ME^{\alpha}{}_M =
\kappa^{\beta}(1+\Gamma)^{\alpha}{}_{\beta}
\end{equation}
where $\kappa^{\beta}(\xi)$ is an anticommuting spacetime spinor but
worldvolume scalar, and where

\begin{equation}
\Gamma^{\alpha}{}_{\beta} = \frac{(-1)^{d(d-3)/4}}{d!{\sqrt
-\gamma}}\epsilon^{i_{1}..i_{d}}E_{i_{1}}{}^{a_{1}}
E_{i_{2}}{}^{a_{2}}\ldots E_{i_{d}}{}^{a_{d}}\Gamma_{a_{1..a{_d}}}~~.
\end{equation}
Here $\Gamma_a$ are the Dirac matrices in spacetime and 
\begin{equation}
\Gamma_{a_{1}..a_{d}}=\Gamma_{[a_1{\cdots}a_{d}]}~~.
\end{equation}
This kappa symmetry has the following important consequences:

\noindent 
1)~~The symmetry is achieved only if certain constraints on the antisymmetric
tensor field strength $F_{MNP..Q}$(Z) and the supertorsion are
satisfied.  In particular the Bianchi identity $dF = 0$ then requires the
$\Gamma$ matrix identity
\begin{equation}
\biggl ( d{\bar \theta} \Gamma_ad\theta\biggr ) \biggl
(d{\bar \theta} \Gamma^{ab_{1}\ldots b_{d-2}} d\theta\biggr )~=0
\end{equation}
for a commuting spinor d$\theta$.  As shown by Achucarro, Evans, Townsend 
and Wiltshire \cite{Achucarro} this is satisfied only for certain values of d
and D.  Specifically, for $d\geq 2$
\begin{eqnarray}
d&=&2:~~~D = 3, 4, 6, 10\nonumber\\
d&=&3:~~~D = 4, 5, 7, 11\nonumber\\
d&=&4:~~~D = 6, 8\nonumber\\
d&=&5:~~~D = 9\nonumber\\
d&=&6:~~~D = 10~~.
\end{eqnarray}
Note that we recover as a special case the well-known result that 
Green-Schwarz superstrings exist {\it classically} only for $D = 3, 4,
6,$ and $10$.  Note also $d_{max} = 6$ and $D_{max} = 11$.  The upper limit
of $D = 11$ is already known in supergravity but there it is necessary to
make extra assumptions concerning the absence of consistent higher spin
interactions.  In supermembrane theory, it follows automatically. See,
however, section (\ref{twelve}).  
~\\

\noindent
2)~~The matrix $\Gamma$ of (1.20) is traceless and satisfies

\begin{equation}
\Gamma^2 = 1
\end{equation}
\noindent
when the equations of motion are satisfied and hence the matrices
$(1\pm\Gamma)/2$ act as projection operators.  The transformation rule
(1.19) therefore permits us to gauge away one half on the fermion degrees of
freedom.  As desribed below, this gives rise to a matching of physical boson
and fermion degrees of freedom on the worldvolume.  
~\\
\noindent
3)~~In the case of the eleven-dimensional supermembrane, it has been shown [16]
that the constraints on the background fields $E_M{}^A$ and $B_{MNP}$
are nothing but the equations of motion of eleven-dimensional
supergravity \cite{Bergshoeff1,Bergshoeff2}.  

\subsection{The brane scan}
\label{scan}

The matching of physical bose and fermi degrees of freedom on the
{\it worldvolume} may, at first sight, seem puzzling since we began with
only spacetime supersymmetry. The explanation is as follows. As the $p$-brane
moves through spacetime, its trajectory is described by the
functions $X^M (\xi)$ where $X^M$ are the spacetime coordinates ($M = 0, 1,
\ldots, D - 1$) and $\xi^i$ are the worldvolume coordinates ($i = 0, 1,
\ldots, d - 1$).  It is often convenient to make the so-called
 {\it static gauge choice} by making the $D = d + (D - d)$ split
\begin{equation}
X^M (\xi) = (X^{\mu} (\xi), Y^m (\xi)),
\end{equation}
where $\mu = 0, 1, \ldots, d - 1$~and~$m = d, \ldots, D - 1$,
and then setting
\begin{equation}
X^{\mu} (\xi) = \xi^{\mu}.
\end{equation}
Thus the only physical worldvolume degrees of freedom are given
 by the $(D - d)~Y^m (\xi)$.  So the number of on-shell bosonic degrees of
freedom is
\begin{equation}
{N_B = D - d.}
\end{equation}
To describe the super $p$-brane we augment the $D$ bosonic coordinates $X^M
(\xi)$ with anticommuting fermionic coordinates $\theta^{\alpha} (\xi)$.
Depending on $D$, this spinor could be Dirac, Weyl, Majorana or
Majorana-Weyl. The fermionic kappa symmetry means that half of the spinor
degrees of freedom are redundant and may be eliminated by a physical gauge
choice.  The net result is that the theory exhibits a {\it $d$-dimensional
worldvolume supersymmetry} \cite{Achucarro} where the number of fermionic
generators is exactly half of the generators in the original spacetime
supersymmetry.  This partial breaking of supersymmetry is a key idea.  Let
$M$ be the number of real components of the minimal spinor and $N$ the
number of supersymmetries in $D$ spacetime dimensions and let $m$~and~$n$
be the corresponding quantities in $d$ worldvolume dimensions.  Let us
first consider $d > 2$.  Since kappa symmetry always halves the number of
fermionic degrees of freedom and going on-shell halves it again, the
number of on-shell fermionic degrees of freedom is
\begin{equation}
{N_F = {1\over 2}mn = {1\over 4}MN.}
\end{equation}
Worldvolume supersymmetry demands $N_B = N_F$ and hence
\begin{equation}
{D - d = {1\over 2}mn = {1\over 4}MN.}
\label{bosefermi}
\end{equation}
A list of dimensions, number of real dimensions of the minimal spinor and
possible supersymmetries is given in Table \ref{minimal}, from which we
see that there are only $8$ solutions of (\ref{bosefermi}) all with $N = 1$, as
shown in Table \ref{branescan}.  We note in particular that $D_{{\rm max}} =
11$ since $M \geq 64$ for $D \geq 12$ and hence (\ref{bosefermi}) cannot be
satisfied.  Similarly
 $d_{{\rm max}} = 6$ since $m \geq 16$ for $d \geq 7$.  The case $d = 2$ is
special because of the ability to treat left and right moving modes
independently.  If we require the sum of both left and right moving bosons
and fermions to be equal, then we again find the condition (\ref{bosefermi}). 
This provides a further $4$ solutions all with $N = 2$, corresponding to
Type $II$ superstrings in $D = 3, 4, 6$~and~$10$ (or 8 solutions
in all if we treat Type $IIA$ and Type $IIB$ separately).  Both the
gauge-fixed Type $IIA$ and Type $IIB$ superstrings will display $(8, 8)$
supersymmetry on the worldsheet. If we require only left (or right) matching,
then  (\ref{bosefermi}) replaced by
\begin{equation}
{D - 2 = n = {1\over 2}MN,}
\end{equation}
which allows another $4$ solutions in $D = 3, 4, 6$~and~$10$,
all with $N = 1$. The gauge-fixed theory will display $(8,0)$ worldsheet
supersymmetry.  The heterotic string falls into this category.  The results
\cite{Achucarro} are indicated by the points labelled $S$ in Table 
\ref{branescan}. Point particles with $d=1$ are usually omitted from the
brane scan \cite{Achucarro,Luscan,Khuristring}, but in Table
\ref{branescan} we have included them.

An equivalent way to arrive at the above conclusions is to list
all scalar supermultiplets and to interpret
the dimension of the target space, $D$, by
\begin{equation}
{D - d =~{\rm number~of~scalars}.}
\label{scalars}
\end{equation}
A useful reference is \cite{Strathdee} which provides an exhaustive
classification of all unitary representations of supersymmetry with maximum
spin $2$.  In particular, we can understand $d_{{\rm max}} = 6$ from this
point of view since this is the upper limit for scalar supermultiplets.
In summary, according to the above classification, Type $II$ $p$-branes do
not exist for $p > 1$. We shall return to this issue, however, in section
(\ref{type}).

\begin{table}
\caption{Minimal spinor components and supersymmetries.}
\label{minimal}
\halign{\indent #&\qquad\hfil# \hfil&\quad\hfil
#\hfil&\quad\hfil # \hfil &
\quad \hfil # \hfil &\quad \hfil # \hfil &\quad #\hfil\cr
&&&Dimension & Minimal Spinor& Supersymmetry&\cr
&&&($D$ or $d$) & ($M$ or $m$) & ($N$ or $n$)&\cr
&&&11 & 32 & 1&\cr
&&&10 & 16 & 2, 1&\cr
&&&9 & 16 & 2, 1&\cr
&&&8 & 16 &2, 1&\cr
&&&7 & 16 & 2, 1&\cr
&&&6 & 8 & 4, 3, 2, 1&\cr
&&&5 & 8 & 4, 3, 2, 1&\cr
&&&4 & 4 & 8, $\ldots$, 1&\cr
&&&3 & 2 & 16, $\ldots$, 1&\cr
&&&2 & 1 & 32, $\ldots$, 1&\cr}
\end{table}

There are four types of solution with  
$8 + 8 $, $4 + 4 $,  $2 + 2 $ or  $1 + 1 $ degrees of freedom respectively. 
Since the numbers  $1 $, $2 $,  $4 $ and  $8 $ are also the dimension of the
four division algebras, these four types of solution are referred to as
real, complex, quaternion and octonion respectively.  The connection with
the division algebras can in fact be made more precise \cite{Sierra,Evans}.

\subsection{String/fivebrane duality in $D=10$?}
\label{five}

Of particular interest was the $D=10$ fivebrane, whose Wess-Zumino term
coupled to a rank six antisymmetric tensor potential $A_{MNPQRS}$ just as the
Wess-Zumino term of the string coupled to a rank two potential $B_{MN}$.
Spacetime supersymmetry therefore demanded that the fivebrane coupled to
the $7$-form field strength formulation of $D=10$ supergravity
\cite{Chamseddine} just as the string coupled to the $3$-form version
\cite{deRoo,Chapline}. These dual formulations of $D=10$ supergravity have long
been something of an enigma from the point of view of superstrings. As field
theories, each seems equally valid. In particular, provided we couple them to
$E_8 \times E_8$ or $SO(32)$ super-Yang-Mills \cite{Greenschwarz}, then
both are anomaly free \cite{Salam,Gates}. Since the $3$-form version
corresponds to the field theory limit of the heterotic string, it was
conjectured \cite{Duffsuper} that there ought to exist a {\it heterotic
fivebrane} which could be viewed as a fundamental anomaly-free theory in
its own right and whose field theory limit corresponds to the dual
$7$-form version. We shall refer to this as the {\it string/fivebrane
duality conjecture}. At this stage, however, the solitonic element had not
yet been introduced.

\subsection{Type $IIA$ superstring in $D=10$ from supermembrane in $D=11$}
\label{reduce}
We begin with the bosonic sector of the $d=3$ worldvolume of the $D=11$
supermembrane:
\begin{eqnarray}
S_3&=&T_3\int d^3\xi\biggl[-{1\over2}\sqrt{-\gamma}\gamma^{ij}
\partial_i X^M\partial_j X^N G_{MN}(X) +{1\over2}\sqrt{-\gamma}\nonumber\\
&&\qquad\qquad
+{1\over3!}\epsilon^{ijk}\partial_i X^M\partial_j X^N\partial_k X^P
C_{MNP}(X)\biggr]\ ,
\label{membrane}
\end{eqnarray}
where $T_3$ is the membrane tension, $\xi^i$ ($i=1,2,3$) are the
worldvolume coordinates, $\gamma^{ij}$ is the worldvolume metric and
$X^M(\xi)$ are the spacetime coordinates $(M=0,1,\ldots,10)$.  Kappa
symmetry \cite{Bergshoeff1,Bergshoeff2} then demands that the
background metric $G_{MN}$ and background 3-form potential $C_{MNP}$
obey the classical field equations of $D=11$ supergravity, whose
bosonic action is
\begin{equation}
I_{11}=\frac{1}{2\kappa_{11}{}^2}\int d^{11}x\sqrt{-G}
\left[R_G-\frac{1}{2\cdot4!}K_{\scriptscriptstyle MNPQ}^2\right]
-\frac{1}{12\kappa_{11}{}^2} \int C_3\wedge K_4 \wedge K_4 \ ,
\label{supergravity11}
\end{equation}
where $K_4=dC_3$ is the 4-form field strength. In particular, $K_4$
obeys the field equation
\begin{equation}
d*K_4=-{1\over2}K_4{}^2
\label{equation4}
\end{equation}
and the Bianchi identity
\begin{equation}
dK_4=0\ .
\label{Bianchi4}
\end{equation}
To see how a double worldvolume/spacetime compactification of the
$D=11$ supermembrane theory on $S^1$ leads to the Type $IIA$ string in
$D=10$ \cite{Howe}, let us denote all $(d=3,D=11)$ quantities by a hat
and all $(d=2,D=10)$ quantities without.  We then make a ten-one split
of the spacetime coordinates
\be
{\hat X}^{\hat M}=(X^M,Y)\qquad M=0,1,\ldots,9
\ee
and a two-one split of the worldvolume coordinates
\begin{equation}
{\hat \xi}^{\hat i}= (\xi^i,\rho)\qquad i=1,2
\end{equation}
in order to make the partial gauge choice
\be
\rho=Y\ ,
\ee
which identifies the eleventh dimension of spacetime with the third
dimension of the worldvolume. The dimensional reduction is then
effected by taking the background fields ${\hat G}_{{\hat
M}{\hat N}}$ and ${\hat C}_{{\hat M}{\hat N}{\hat P}}$ to be independent of
$Y$.  The string backgrounds of dilaton $\Phi$, string $\sigma$-model metric
$G_{MN}$, $1$-form $A_M$, $2$-form $B_{MN}$ and $3$-form $C_{MNP}$ are given
by
\footnote{The choice of dilaton prefactor, $e^{-\Phi/3}$, is dictated by the
requirement that $G_{MN}$ be the $D=10$ string $\sigma$-model metric.  To
obtain the $D=10$ fivebrane $\sigma$-model metric, the prefactor is unity
because the reduction is then spacetime only and not simultaneous
worldvolume/spacetime.  This explains the remarkable ``coincidence''
\cite{Lublack} between $\hat G_{MN}$ and the $D=10$ fivebrane $\sigma$-model
metric.}
\begin{eqnarray}
{\hat G}_{MN}&=& e^{-\Phi/3}\left(
\begin{array}{cc}
G_{MN}+e^\Phi A_MA_N&e^{\Phi}A_M\\
e^{\Phi}A_N&e^{\Phi}
\end{array}
\right)\nonumber\\
{\hat C}_{MNP}&=&C_{MNP}\nonumber\\
{\hat C}_{MNY}&=&B_{MN}\ .
\end{eqnarray}

The actions (\ref{membrane}) and (\ref{supergravity11}) now reduce to
\begin{eqnarray}
S_2=T_2\int d^2\xi\biggl[&-&{1\over2}\sqrt{-\gamma}\gamma^{ij}
\partial_i X^M\partial_j X^N G_{MN}(X) 
\nonumber \\
&-& {1\over2!}\epsilon^{ij}\partial_i X^M\partial_j X^N
B_{MN}(X)+\cdots\biggr] 
\end{eqnarray}
and
\begin{eqnarray}
I_{10}&=&\frac{1}{2\kappa_{10}{}^2}\int d^{10}x\sqrt{-G}e^{-\Phi} \Big[
R_G+(\partial_{\scriptscriptstyle M}\Phi)^2
-\frac{1}{2\cdot3!}H_{\scriptscriptstyle MNP}^2
-\frac{1}{2\cdot2!}e^\Phi F_{\scriptscriptstyle MN}^2
 \nonumber\\
&&~~~~~~~~~~-\frac{1}{2\cdot4!}e^\Phi J_{\scriptscriptstyle MNPQ}^2 \Big]
-{1\over2\kappa_{10}{}^2}\int{1\over2}K_4\wedge K_4\wedge B_2\ ,
\label{eq:supergravity10}
\end{eqnarray}
where the field strengths are given by $J_4=K_4+A_1H_3$, $H_3=dB_2$ and
$F_2=dA_1$.

One may repeat the procedure in superspace to obtain
\begin{eqnarray}
S_2=T_2\int
d^2\xi\biggl[-{1\over2}\sqrt{-\gamma}\gamma^{ij}{E_i{}^aE_j{}^b\eta_{ab}} 
+{1\over2!}\epsilon^{ij}\partial_i X^M\partial_j X^N
B_{MN}(Z)\biggr] 
\label{string}
\end{eqnarray}
which is just the action of the Type $IIA$ superstring.

The double dimensional reduction described above seems superficially to be
equivalent to considering an $R^{10} \times S^1$ spacetime, taking $Y$ to be
the coordinate on the circle and shrinking the radius to zero.
However, it has recently been pointed out \cite{Russo} that this latter
procedure still retains the zero modes responsible for the membrane
instability discussed in section (\ref{massless}). These membrane
instabilities are absent when the membrane is wrapped around the
torus of an $R^{9} \times S^1 \times S^1$ spacetime
\cite{Inami,Russo,Russotseytlin}. Indeed, such a wrapped membrane
exactly reproduces the $(q_1,q_2)$ bound states of the $NS-NS$ and $R-R$
strings in Type $IIB$ theory \cite{Schwarzpower}. 

\subsection{The signature of spacetime}
\label{signature}

If our senses are to be trusted, we live in a world with three space and
one time dimensions.  However, the revival of the Kaluza-Klein idea
\cite{Kaluza,Klein}, brought about by supergravity and superstrings, has
warned us that this may be only an illusion.  In any case, there is a
hope, so far unfulfilled, that the four-dimensional structure that we
apparently observe may actually be predicted by a {\it Theory of
Everything}.  Whatever the outcome, imagining a world with an arbitrary
number of space dimensions has certainly taught us a good deal about the
properties of our three-space-dimensional world.

In spite of all this activity, and in spite of the popularity of Euclidean
formulations of field theory, relatively little effort has been devoted to
imagining a work with more than one {\it time} dimension.  This
is no doubt due partially to the psychological difficulties we have in
treating space and time on the same footing.  As H. G. Wells reminds us in
{\it The Time Machine}: ``There is, however, a tendency to draw an unreal
distinction between the former three dimensions and the latter, because it
happens that our consciousness moves intermittently in one direction along
the latter from the beginning to the end of our lives."  There are also
more justifiable reasons associated with causality.  Nevertheless, one
might hope that a theory of everything should predict no only the
{\it dimensionality} of spacetime, but also its {\it signature}.

For example, quantum consistency of the superstring requires $10$ spacetime
dimensions, but not necessarily the usual $(9, 1)$ signature.  The
signature is not completely arbitrary, however, since spacetime
supersymmetry allows only $(9,1)$, $(5,5)$ or $(1,9)$.  Unfortunately,
superstrings have as yet no answer to the question of why our universe
appears to be four-dimensional, let alone why it appears to have
signature $(3,1)$.

The authors of \cite{Blencowe} therefore considered a world with an
arbitrary number $T$ of time dimensions and an arbitrary number $S$ of space
dimensions to see how far classical supermembranes restrict not only $S + T$
but $S$ and $T$ separately.  To this end they also allowed an $(s,t)$ signature
for the worldvolume of the membrane where $s \leq S$ and $t \leq T$ but are
otherwise arbitrary.  It is not difficult to repeat the analysis of section
(\ref{scan}) for arbitrary signatures, and to show that there is once again a
matching of the bosonic and fermionic degrees of freedom as a consequence of
the kappa symmetry.  However severe constraints on possible supermembrane
theories will now follow by demanding spacetime supersymmetry.

We restrict ourselves in this section to $N = 1$ flat
superspace, and require $\theta^{\alpha}$ to be a minimal
spinor i.e. Majorana, Weyl, Majorana-Weyl etc., whenever it is possible to
impose such a condition.  We furthermore assume invariance under the
generalized super-poincare group super-$IO(S,T)$, as required by the
superspace construction of section (\ref{scan}).  Such supermembranes (and
their compactifications) are the only ones currently known.  In section
(\ref{twelve}) we shall consider the possibility of other supermembrane
theories obtained by requiring spacetime supersymmetry but with a different
supergroup.  For the moment, however, we shall require super-Poincare which
means, in particular, that the anticommutator of two supersymmetry 
charges $Q$ yields a translation
\begin{equation}
\{Q, Q\} \sim P~~.
\end{equation}
This is only possible for certain values of $S$ and $T$ when $Q$ is a
minimal spinor. 

Those values of $S$ and $T$ permitting minimal spinors have been determined
by Kugo and Townsend \cite{Kugo}.  See also the works of van Nieuwenhuizen
\cite{vanNieuwenhuizen} , Coqueraux \cite{Coqueraux} and Freund
\cite{Freund}.  For the Clifford algebra given by
$\Gamma^a\Gamma^b+\Gamma^b\Gamma^a= 2\eta^{ab}$ there exist matrices $A$ and
$B$ for which
\begin{equation}
\Gamma_a^{\dagger}=(-1)^T A\Gamma_a A^{-1},~~AA^{\dagger} = 1
\end{equation}
\begin{eqnarray}
\Gamma_a=\eta  B^{-1} \Gamma_a^{\ast}B,~~&BB^{\dagger}=1,\nonumber\\
&B^{\ast}B = \epsilon
\end{eqnarray}
where $\epsilon$ and $\eta$ are given in Table \ref{epsilon}.
\begin{table}
\begin{center}
\begin{tabular}{ccc}
$S-T$ mod $8$ & $\epsilon$ & $\eta$\\ 
~\\
$0,1,2$ & $+1$ & $+1$\\
$6,7,8$ & $+1$ &$ -1$\\
$4,5,6$ & $-1$ & $+1$\\
$2,3,4$ & $-1$ & $-1$
\end{tabular}
\end{center}
\caption{Values of $\epsilon$ and $\eta$}
\label{epsilon}
\end{table}
We can choose a basis such that

\begin{eqnarray}
\Gamma_a^{\dagger} = &-&\Gamma_a~~a = 1,\ldots,T\nonumber\\
&+&\Gamma_a~~a=T+1,\ldots D
\end{eqnarray}
and
\begin{equation}
A = \Gamma_1\Gamma_2\ldots\Gamma_T~~.
\end{equation}
The charge conjugation matrix is defined as $C = {\tilde B}A$.  The
properties of $A$ and $ B$ then imply
\begin{eqnarray}
{\tilde\Gamma} &=& (-1)^T\eta C\Gamma C^{-1}\\
{\tilde C} &=& \epsilon\eta^T(-1)^{T(T+1)/2}C,~~C^{\dagger}C = 1
\end{eqnarray}
where the tilde  denotes transpose.  For $D$ even, we can also define the
projection operator
\begin{equation}
P_{\pm}= \frac{1}{2}\biggl [ 1\pm(-1)^{(S-T)/4} \Gamma^{D+1}\biggr
] 
\end{equation}
where
\begin{equation}
\Gamma^{D+1} = \Gamma^1\Gamma^2\ldots\Gamma^D~.
\end{equation}
Using the above properties we find that we can have the minimal
spinors given in Table \ref{spinor}.

The next task is to check which of these possibilities admits
the super-poincare algebra.  The part of the superalgebra
which is the same in each case is
\[
[M_{ab}, M_{cd}]= -i
(\eta_{bc} M_{ad} -\eta_{ac} M_{bd}-\eta_{bd} M_{ac}+\eta_{ad} M_{bc})
\]
\[
[M_{ab}, P_c]= i (\eta_{ac} P_b-\eta_{bc} P_a)
\]
\[
[P_a, P_b]= 0
\]
\begin{equation}
[M_{ab}, Q_{\alpha}]= -\frac{i}{2} (\Gamma_{ab}Q)_\alpha~~.
\end{equation}
We now examine the \{Q,Q\} anticommutator.  Consider first $S - T = 0, 1,
2$ mod $8$ for which $Q_\alpha$ is Majorana.  The only possible form for
the anticommutator is
\begin{equation}
\{Q_{\alpha}, Q_{\beta}\} = (\Gamma^a C^{-1})_{\alpha\beta}P_a.
\end{equation}
Since the left hand side is symmetric under interchange of $\alpha$ and
$\beta$ we require
\begin{equation}
{\tilde {(\Gamma_a C^{-1})}} = \Gamma_a C^{-1}
\end{equation}
but
\begin{eqnarray}
(\Gamma_a C^{-1})&=&{\tilde C}^{-1}{\tilde\Gamma}_a\nonumber\\
&=&\epsilon\eta^T (-1)^{T(T+1)/2}C^{-1}{\tilde\Gamma}_a\nonumber\\
&=&\epsilon\eta^{T+1}(-1)^{T(T-1)/2}\Gamma_a C^{-1}~~.
\end{eqnarray}
Now from Table \ref{epsilon}, $\epsilon = \eta + 1 $ for $S - T = 0,1,2$
mod $8$ and hence 
\begin{equation}
{\tilde {(\Gamma_a C^{-1})}} = (-1)^{T(T-1)/2}\Gamma_a C^{-1}
\end{equation}
which is compatible with (1.43) only if $T = 0,1$ mod $4$.  Now consider
the subcase $S - T= 0$ mod $8$, $T = 0, 1$ mod $4.$  Define
\begin{equation}
Q_{\pm \alpha}=(P_{\pm}Q)_{\alpha}.
\end{equation}
From (1.42) we have
\begin{eqnarray}
\{Q_{\pm\alpha}, Q_{\pm\beta}\}&=&P_{\pm\alpha\gamma}
\{Q_{\gamma}, Q_{\beta}\} P_{\pm\delta\beta}\nonumber\\
&=&(P_{\pm}\Gamma_a C^{-1}{\tilde P}_{\pm})_{\alpha\beta}P_a
\end{eqnarray}
but
\begin{eqnarray}
C^{-1}{\tilde
P}_{\pm}&=&C^{-1}\biggl(1\pm{\tilde\Gamma}_D{\tilde\Gamma}_{D-1}
\ldots{\tilde\Gamma}_1\biggr)C^{-1}\nonumber\\
&=&\biggl(1\pm\Gamma_D\Gamma_{D-1}\cdots\Gamma_1\biggr)\nonumber\\
&=&\biggl(1\pm(-1)^{D(D-1)/2}\Gamma^{D+1}\biggr) C^{-1}
\end{eqnarray}
and therefore
\begin{eqnarray}
P_{\pm}\Gamma_a C^{-1}{\tilde
P}_{\pm}&=&P_{\pm}\biggl(1\mp(-1)^{T(2T-1)}\Gamma^{D+1}\biggr)\Gamma_a
C^{-1}\nonumber\\ 
&=&\left\{\begin{array}{ll}
  0 & \mbox{T =0 mod 4}\\ 
P_{\pm}\Gamma_a C^{-1} &\mbox{T = 1 mod 4~~.}\nonumber
\end{array}
\right. 
\end{eqnarray}
Thus splitting up $\{Q,Q\} = \Gamma^aC^{-1}P_a$ into its chiral
parts, we get
\begin{eqnarray}
\{Q_{\pm},Q_{\pm}\} &=& P_{\pm}\Gamma^aC^{-1}P_a\nonumber\\
\{Q_{\pm},Q_{\mp}\} &=& 0
\end{eqnarray}
for $T = 1$ mod $4$ and
\begin{eqnarray}
\{Q_{\pm},Q_{\pm}\} &=& 0\nonumber\\
\{Q_{\pm},Q_{\mp}\} &=& P_{\pm}\Gamma^a C^{-1}P_a
\end{eqnarray}
for $T = 0$ mod $ 4$.  Thus for $ S - T = 0$ mod $8$, only for $T = 1$ mod $4$
can we set $Q_- = 0$ say, and obtain $\{Q,Q\} \sim P$ with $Q$ a
Majorana-Weyl minimal spinor.

We can proceed in this way to exhaust all the possible values of $S$ and $T$
admitting super-Poincare symmetry.  These are summarized in Table
\ref{spinor}, where ${\bar Q}=Q^{\dagger}A$.
\begin{table}
\caption{Minimal spinors for different values of $S - T$ mod $8$ and
super-Poincare algebras for different values of $T$ mod $4$.}
\label{spinor}
\begin{tabular}{llll}
S - T mod 8 & Minimal spinor type & T mod 4 & Anticommutator\\
~&~&~&~\\
1,2 & Majorana & 0,1 & $\{Q,Q\} = \Gamma^a C^{-1}P_a$\\
6,7 & pseudo-Majorana & 1,2 & $\{Q,Q\}  = i\Gamma^aC^{-1}P_a$\\
0 & Majorana-Weyl $\rbrace$ & 1 &$\{Q_+,Q_+\} = P_+\Gamma^aC^{-1}P_a$\\
~& pseudo-Maj-Weyl $\rbrace$ &  ~  & $\{Q_+,Q_+\} = iP_+\Gamma^aC^{-1}P_a$\\
3,5 & Dirac & 0,1 & $\{Q,\bar Q\} =\Gamma^a P_a$\\
~& & 2,3 & $\{Q,\bar Q\} = i\Gamma^a P_a$\\
~& & ~ ~ & $\{ Q,  Q\}=\{\bar Q, \bar Q\} = 0$\\
4 & Weyl & 1 & $\{Q_+, \bar Q_+\} = P_+ \Gamma^a P_a$
\end{tabular}
\end{table}
Combining these results with those of section (\ref{scan}) allows to draw the
$S/T$ plot of Table \ref{molecule} whose points correspond to possible
supermembrane theories.  Once again we have used the symbols $O, H, C$ and $
R$ to denote objects with $8+8$, $4+4$, $2+2$, or $1+1$ degrees of freedom,
respectively. The points marked $\tilde O,\tilde  H,\tilde  C$ and $
\tilde R$ are not permitted if, as we have assumed in this section, the
algebra is super-Poincare. For pictorial reasons, we call this table the
{\it brane-molecule}. 
\begin{table}
\caption{ The brane-molecule (for scalar multiplets with $d \geq 2$ only),
where $R$, $C$, $H$ and $O$ denote real ($1+1$), complex ($2+2$), quaternion
($4+4$) and octonion ($8+8$), respectively. Those marked with a tilde, in
particular the $(2,2)$-brane in $(10,2)$, are forbidden if the algebra is
super-Poincare.}
\label{molecule}
$
\begin{array}{cccccccccccccc}
S\uparrow&&&&&&&&&&&~\\
11&.&&&&&&&&&&&~\\
10&.&O&\tilde O&&&&&&&&&~\\
9&H&H/O&O&&&&&&&&&~\\
8&\tilde C&H&&&&&&&&&&~\\
7&\tilde C&H&&&&&&&&&&~\\
6&\tilde C/\tilde H&H&&&&O&\tilde O&&&&~\\
5&C&C/H&H&H&H&H/O&O&&&&&~\\
4&\tilde C&C&\tilde C&\tilde C&\tilde C&H&&&&&&~\\
3&.&R/C&\tilde R&\tilde R&\tilde C&H&&&&&&~\\
2&.&R&~&\tilde R&\tilde C&H&~&~&~&O&\tilde O&~\\
1&.&~&R&R/C&C&C/H&H&H&H&H/O&O&~\\
0&.&.&.&.&\tilde C&C&\tilde C/\tilde H&\tilde C&\tilde C&H&.&.~\\
~&0&1&2&3&4&5&6&7&8&9&10&11&T\rightarrow
\end{array}
$
\end{table}

Several comments are now in order:

\noindent
1)~~In the absence of any physical boundary conditions which treat time
differently from space, and which we have not yet imposed, the mathematics
will be symmetric under interchange of $S$ and $T$.  This can easily be seen
from Table \ref{molecule}.  For every supermembrane with $(S,T)$
signature, there is  another with $(T,S)$.  Note the self-conjugate theories
that lie on the  $S = T$ line which passes through the $(5,5)$ superstring.

\noindent
2)~~There is, as yet, no restriction on the worldvolume signatures beyond
the original requirement that $s\leq S$ and $t\leq T$.

\noindent
3)~~If we were to redraw the $D/d$ brane scan of Table \ref{branescan}
allowing now arbitrary signature, there would be no new $S$ points on the
plot, but rather the new solutions would be superimposed on the old ones. 
For example, there would now be six solutions occupying the $(d = 3, D =
11)$ slot instead of one. 

\noindent
4)~~Perhaps the most interesting aspect of the brane-molecule is the mod $8$
periodicity.  Suppose there exist signatures $(s,t)$ and $(S,T)$ which satisfy
both the requirements of bose-fermi matching and super-Poincare invariance. 
Now consider $(s',t')$ and $(S',T')$ for which
\begin{equation}
s'+t' = s +t
\end{equation}
\begin{equation}
S'+T' = S+T~~.
\end{equation}
As a consequence of the modulo $8$ periodicity theorem for real Clifford
algebras, the minimal condition on a spinor is modulo $8$ periodic 
e.g. $S - T =0$ mod $8$ for Majorana-Weyl.  So if, in addition to (1.51) and
(1.52) we also have
\begin{equation}
S'-T'=S-T+8n~~~~n\epsilon Z~~~.
\end{equation}
then $(s',t')$ and $(S',T')$ satisfy bose-fermi matching.  (1.52) and
(1.53) imply
\begin{eqnarray}
S'&=& S + 4n\\
T'&=&T - 4n~~~.
\end{eqnarray}
Since, from Table \ref{spinor}, the existence of a super-Poincare algebra
with minimal spinors is modulo $4$ periodic in $T$, we see from (1.54) that 
the
super-Poincare invariance is also satisfied.  Thus given the vertical
sequence $(S,T) = (10,1) \rightarrow (2,1)$, modulo $ 8$ periodicity implies
the existence of the two other vertical sequences of Table \ref{molecule},
namely $(6,5) \rightarrow (0,5)$ and $(2,9) \rightarrow (0,9)$.  The three
horizontal sequences $(1,10) \rightarrow (1,2), (5,6) \rightarrow (5,0)$ and
$(9,2) \rightarrow (9,0)$ are similarly related via modulo $8$ periodicity.

Note the special crossover points at $(9,1), (5,5)$ and $(1,9)$ which permit
Majorana-Weyl spinors and which correspond to the top horizontal line in the
brane scan of Table \ref{branescan}.  Similarly Weyl spinors are permitted
at  the
crossover points $(5,1)$ and $(1,5)$ corresponding to the middle horizontal
line of Table \ref{branescan}.  It is curious that the fundamental extended
objects at the top  of the $H$ and $C$ sequences are chiral, while those at
the top of the $O$ and $R$ sequences are not. 

In the usual signature all extended objects appear to suffer from ghosts
because the kinetic term for the $X^0$ coordinate enters with the wrong
sign.  These are easily removed, however (at least at the classical level) 
by the presence of diffeomorphisms on the worldvolume which allow us to
fix a gauge where only positive-norm states propagate e.g. the light cone
gauge for strings and its membrane analogues. See section
(\ref{lightcone}).  Alternatively we may identify the $d$ worldvolume
coordinates $\xi^i$ with $d$ of the $D$ space-time coordinates $X^i$ $(i =
1,2,3)$ leaving $(D -d)$ coordinates $X^I$ $(I=1..D-d)$ with the right
sign for their kinetic energy \cite{Bergshoeffgamma}.  Of course, this
only works if we have one worldvolume time coordinate $\tau$ that allows
us to choose a light-cone gauge or else set $\tau = t$.

In the same spirit, we could now require absence of ghosts (or rather
absence of classical instabilities since we are still at the classical
level) for arbitrary signature by requiring that the ``transverse" group
$SO(S - s, T - t)$ which governs physical propagation after gauge-fixing, be
compact.  This requires $T = t$.

It may be argued, of course, that in a world with more than one time
dimension, ghosts are the least of your problems.  Moreover, in contrast to
strings, unitarity on the worldvolume does not necessarily imply  unitarity
in spacetime (I am grateful for discussions on this point with J.
Polchinski).  This is becuase the transverse group no longer coincides with
the little group.  (For example, the $(2,1)$ object in $(10,1)$ spacetime 
and the $(1,2)$ object in $(9,2)$ spacetime both have transverse group
$SO(8)$, but the former has little group $SO(9)$ and the latter $ SO(8,1)$.) 
Nevertheless, it is an interesting exercise to see how compactness of the
gauge group restricts the possible super-extended-objects.  For example, the
superstring in $(9,1)$ survives with $SO(8)$, but the superstring in $(5,5)$
with $SO(4,4)$ does not.  What about the superstring in $(1,9)$?  Here we
once again encounter the problem that, in the absence of any physical input,
we cannot distinguish $(S,T)$ signature from $(T,S)$.  Since positivity of
the energy is only a convention in field theory, ghosts can still be avoided
by choosing $S = s$ instead of $T = t$.  The best thing is simply to
cut the Gordian knot and demand that $S\geq T$. The resulting
restricted brane molecule may be found in \cite{Blencowe}. 

\subsection{Twelve dimensions?}
\label{twelve}

It is interesting to ask whether we have exhausted all possible theories of
extended objects with spacetime supersymmetry and fermionic gauge invariance
on the worldvolume and fields $(X^M,\theta^\alpha)$.  This we claimed to
have done in section (\ref{signature}) by demanding super-Poincare
invariance but might there exist other Green-Schwarz type actions in which
the supergroup is not necessarily super-Poincare?  Although we have not yet
attempted to construct such actions, one may nevertheless place
constraints on the dimensions and signatures for which such theories are
possible.  We simply impose the bose-fermi matching constraints of section
(\ref{scan}) and those in the first two columns of Table \ref{spinor} but
relax the constraints  in the second two columns which specifically
assumed super-Poincare invariance.  The extra allowed theories
\cite{Blencowe} are indicated by the points marked $\tilde R,\tilde 
C,\tilde  H$ and $\tilde  O$ in Table \ref{molecule}.

Although the possibilities are richer,
there are still several constraints.  Note in particular that the maximum
space-time dimension is now $D=12$ provided we have signatures $(10,2)$,
$(6,6)$ or $(2,10)$.  These new cases are particularly
interesting since they belong to the $O$ sequence and furthermore admit
Majorana-Weyl spinors.  In fact, the idea of a twelfth
timelike dimension in supergravity is an old one
\cite{Julia} and twelve-dimensional supersymmetry algebras have been
discussed in the supergravity literature \cite{vanHolten}.  In particular,
the chiral $(N_+,N_-)=(1,0)$ supersymmetry algebra in $(S,T)=(10,2)$
involves the anti-commutator    \be
\{Q_{\alpha},Q_{\beta}\}=\Gamma^{MN}{}_{\alpha \beta}P_{MN}
+\Gamma^{MNPQRS}{}_{\alpha \beta}Z^+{}_{MNPQRS} \ee
The right hand side yields not only a Lorentz generator but also a six index
object so it is certainly not super-Poincare.  

In section (\ref{reduce}), we obtained the Type $IIA$ string
by compactifying the $D=11$ supermembrane on a circle.  An obvious question,
therefore, is whether Type $IIB$ also admits a  higher-dimensional
explanation.  It was conjectured in \cite{Blencowe} that the $(2,2)$ extended
object moving in $(10,2)$ space-time may (if it exists) be related by
simultaneous dimensional reduction to the $(1,1)$ Type $IIB$ superstring in
$(9,1)$ just as the $(2,1)$ supermembrane in $(10,1)$ is related to the Type
$IIA$ superstring .  Evidence in favor of the conjecture was supplied by the
appearance of Majorana-Weyl spinors and self-dual tensors in both the
twelve-dimensional and Type $IIB$ theories. This idea becomes even more
appealing if one imagines that the $SL(2,Z)$ of the Type $IIB$ theory
\cite{Hulltownsend} might correspond to the modular group of a $T^2$
compactification from $D=12$ to $D=10$ just as the $SL(2,Z)$ of $S$-duality
corresponds to the modular group of a $T^2$ compactification from $D=6$ to
$D=4$ \cite{Duffstrong}.  On the other hand, the absence of translations
casts doubt on the naive application of the bose-fermi matching argument,
and the appearance of the self-dual $6$-form charge $Z$ is suggestive of a
sixbrane, rather than a threebrane. 

Despite all the objections one might raise to a world with two time
dimensions, and despite the above problems of interpretation, the idea of a
$(2,2)$ object moving in a $(10,2)$ spacetime has recently been revived in
the context of {\it $F$-theory} \cite{Vafa}, which involves Type $IIB$
compactification where the axion and dilaton from the Ramond-Ramond sector
are allowed to vary on the internal manifold.  Given a manifold $M$ that has
the structure of a fiber bundle whose fiber is $T^2$ and whose base is some
manifold $B$, then  \be
F~ on ~M \equiv ~Type~IIB~on~B      
\ee
The utility of $F$-theory is beyond dispute and it has certainly enhanced
our understanding of string dualities, but should the twelve-dimensions
of $F$-theory be taken seriously? And if so, should  $F$-theory be
regarded as more fundamental than $M$-theory? Given that there seems to be
no supersymmetric field theory with $SO(10,2)$ Lorentz
invariance \cite{Nishino}, and given that the on-shell states carry only
ten-dimensional momenta \cite{Vafa}, the more conservative interpretation
is that the twelfth dimension is merely a mathematical artifact and that
$F$-theory should simply be interpreted as a clever way of compactifying
the $IIB$ string \cite{Sen}.  Time ( or should I say ``Both times''?) will
tell.   

\subsection{Type $II$ $p$-branes: the brane scan revisited}
\label{type}

According to the classification of \cite{Achucarro} described in
section (\ref{scan}), no Type $II$ $p$-branes with $p > 1$ could exist.
Moreover, the only brane allowed in $D=11$ was $p=2$. These  conclusions
were based on the assumption that the only fields propagating on the
worldvolume were scalars and spinors, so that, after gauge fixing, they fall
only into {\it scalar} supermultiplets, denoted by $S$ on the brane scan of
Table \ref{branescan}. Indeed, these were the only kappa symmetric actions
known at the time. Using soliton arguments, however, it was pointed out in
\cite{Callan1,Callan2} that both Type $IIA$ and Type $IIB$ superfivebranes
exist after all. Moreover, the Type $IIB$ theory also admits a self-dual
superthreebrane \cite{Luthree}. The no-go theorem is circumvented because in
addition to the superspace coordinates $X^M$ and $\theta^\alpha$ there are
also higher spin fields on the worldvolume: vectors or antisymmetric
tensors. This raised the question: are there other super $p$-branes and if
so, for what $p$ and $D$? In \cite{Luscan} an attempt was made to answer
this question by asking what new points on the brane scan are permitted by
bose-fermi matching alone. Given that the gauge-fixed theories display
worldvolume supersymmetry, and given that we now wish to include the
possibility of vector and antisymmetric tensor fields, it is a
relatively straightforward exercise to repeat the bose-fermi matching
conditions of the section (\ref{scan}) for vector and antisymmetric
tensor supermultiplets.  

Let us begin with vector supermultiplets. Once again, we may proceed in one 
of two ways. First, given that a worldvolume vector has ($d - 2$) degrees of
freedom, the scalar multiplet condition (\ref{bosefermi}) gets replaced by 
\begin{equation} 
{D - 2 = {1\over 2}~mn ={1\over 4}~MN .}
\end{equation}
Alternatively, we may simply list all the vector supermultiplets in the
classification of \cite{Strathdee} and once again interpret $D$ via
(\ref{scalars}). The results \cite{Luscan,Khuristring} are shown by the
points labelled $V$ in Table \ref{branescan}.  

Next we turn to antisymmetric tensor multiplets. In $d = 6$ there is a
supermultiplet with a second rank tensor whose field strength is self-dual: 
$(B_{\mu\nu}^-, \lambda^I, \phi^{[IJ]})$, $I = 1, \ldots, 4$.  This is has
chiral $d=6$ supersymmetry. Since there are five scalars, we have
$D=6+5=11$. There is thus a new point on the scan corresponding to the $D=11$
superfivebrane.  One may decompose this $(n_+,n_-)=(2,0)$ supermultiplet
under $(n_+,n_-)=(1,0)$ into a tensor multiplet with one scalar and a
hypermultiplet with four scalars.  Truncating to just the tensor multiplet
gives the zero modes of a fivebrane in $D=6+1=7$. These two tensor multiplets
are shown by the points labelled $T$ in Table \ref{branescan}.
\footnote{I am grateful to Paul Townsend for pointing out that the
$D=11$ fivebrane was inadvertently omitted from the brane scan
in \cite{Luscan} and to Ergin Sezgin for pointing out that
the the $D=7$ fivebrane was inadvertently ommitted from the brane scan in
\cite{Khuristring}.} 

Several comments are now in order:

1) The number of scalars in a vector supermultiplet is such that, from
(\ref{scalars}), $D = 3, 4, 6$ or $10$ only, in accordance with
\cite{Strathdee}.

2) Vector supermultiplets exist for all $d \leq10$ \cite{Strathdee}, as may
be seen by dimensionally reducing the $(n=1,d=10)$  Maxwell
supermultiplet.  However, in $d = 2$ vectors have no degrees of freedom and
in $d = 3$ vectors have only one degree of freedom and are dual to scalars. 
In this sense, therefore, these multiplets will already have been included as
scalar multiplets in section (\ref{scan}). There is consequently some
arbitrariness in whether we count these as new points on the scan and in
\cite{Luscan,Khuristring} they were omitted.  For example, it was recognized
\cite{Luscan} that by dualizing a vector into a scalar on the gauge-fixed
$d=3$ worldvolume of the Type $IIA$ supermembrane, one increases the number
of worldvolume scalars, {\it i.e.} transverse dimensions, from $7$ to $8$
and hence obtains the corresponding worldvolume action of the $D=11$
supermembrane.  Thus the $D=10$ Type $IIA$ theory contains a hidden $D=11$
Lorentz invariance \cite{Luscan,Schmidhuber,TownsendD}!

3) This dualizing of the scalar into a vector on the $3$-dimensional
worldvolume, which has the effect of lowering the spacetime
dimension by one, is a special case of a more general phenomenon of dualizing
scalars into antisymmetric tensors of rank $(d-2)$ on a $d$-dimensional
worldvolumes. For example in \cite{Howesezgin}, it is argued that one
should also include new points on the scan with ($d=6,D=9$), 
($d=5,D=8$),  ($d=4,D=7$) and  ($d=4,D=5$) obtained by dualizing one of the
four scalars in the hypermultiplets describing the known points at
($d=6,D=10$),  ($d=5,D=9$),   ($d=4,D=8$) and  ($d=4,D=6$). The problem
with this is knowing when to stop. Although \cite{Strathdee} lists the
particle content in all the multiplets allowed by supersymmetry, it does
not list all possible field representations. Depending on its
interactions, a scalar may or may not admit a dualization. An interesting
example of this dilemma is provided by the fivebrane multiplet
$(B_{\mu\nu}^-, \lambda^I, \phi^{[IJ]})$,  $I = 1, \ldots, 4$ which has
five scalars and hence seems to be intrinsically eleven dimensional.
However, it was originally proposed as the worldvolume of the $D=10$ Type
$IIA$ fivebrane \cite{Callan1,Callan2}. Does this interpretation therefore
demand that we can dualize one of the scalars? Or should we turn the
argument around to say that the inability to dualize is yet another
indication that the strong coupling limit of the Type $IIA$ theory is
necessarily eleven-dimensional? 

4) In listing vector multiplets, we have focussed only on the abelian
theories obtained by dimensionally reducing the Maxwell multiplet. One
might ask what role, if any, is played by non-abelian Yang-Mills
multiplets. See below. 

5) We emphasize that the points labelled $V$ and $T$ merely tell us what is
allowed by bose/fermi matching. We must then try to establish which of these
possibilities actually exists. When this scan was first written down in 1993
\cite{Luscan} we knew of the following soliton solutions:  First the Type
$IIA$ and Type $IIB$ superfivebranes \cite{Callan1,Callan2} and the
self-dual Type $IIB$ superthreebrane \cite{Luthree} (the first example of a
supermembrane carrying Ramond-Ramond charges) all found in 1991, then the
$D=11$ superfivebrane \cite{Gueven} found in 1992, then the Type $IIA$ $p$
branes with all $p=0,1,2,3,4,5,6$ found in 1993. The other points labelled
$V$ were still something of a mystery. To see why these choices of $p$ were
singled out, we recall that Type $II$ string theories differ from heterotic
theories in one important respect: in addition to the usual Neveu-Schwarz 
charge associated with the $3$-form field strength, they also carry 
Ramond-Ramond charges associated with $2$-form and $4$-form field strengths in
the case of Type $IIA$ and $3$-forms and $5$-forms in the case of
Type $IIB$. Accordingly, the new solutions of the Type $IIA$ string equations
were found to describe {\it  electric} super $p$-branes with $p=0,2$ and
their {\it magnetic} duals with $p=6,4$ and new solutions of Type $IIB$
string equations were found to describe {\it electric} super $p$-branes with
$p=1,3$ and their {\it magnetic} duals with $p=5,3$.  Interestingly enough,
the Type $IIB$ superthreebrane is {\it self-dual}, carrying a magnetic charge
equal to its electric charge \cite{Luthree}.

However, the whole subject of Type $II$ supermembranes underwent a major sea
change in 1995 when Polchinski \cite{Polchinski} realized that
Type $II$ super $p$-branes carrying Ramond-Ramond charges admit the
interpretation of {\it Dirichlet-branes} that had been proposed earlier in
1989 \cite{Dai}. These $D$-branes are surfaces of dimension $p$ on which open
strings can end. The Dirichlet $p$-brane is defined by Neumann boundary 
conditions in $(p+1)$ directions (the worldvolume itself) and Dirichlet
boundary conditions in the remaining $(D-p-1)$ transverse directions. In
$D=10$, they exist for even $p=0,2,4,6,8$ in the Type $IIA$ theory and
odd $p=-1,1,3,5,7,9$  in the Type $IIB$ theory, in complete
correspondence with the points marked $V$ on the brane scan of Table
\ref{branescan}. The fact that these points preserve one half of the
spacetime supersymmetry and are described by dimensionally reducing the
$(n=1,d=10)$ Maxwell multiplet fits in perfectly with the $D$-brane picture.
In hindsight there also exists a Type $IIB$ supergravity interpretation of
the $(-1)$-brane, which is an instanton, and its $7$-brane dual
\cite{Gibbons}. The $9$-brane emerges from the fact that purely Neumann
strings can end anywhere.  Also in hindsight the $8$-branes of the Type $IIA$
theory also admit a soliton interpretation \cite{Bergshoeff4} because there
exists a version of $D=10$ Type $IIA$ supergravity with a gravitino mass
term and cosmological constant proposed by Romans \cite{Romans}. But in $D$
dimensions a cosmological term is equivalent to a rank $D$ antisymmetric
tensor field strength \cite{Duffvan,Aurelia}, and hence yields a $(D-2)$
-brane, which is a {\it domain wall}. Domain walls are dual to
$(-2)$-branes \cite{Khurinew} (whatever that means). 

Moreover when $N$ branes coincide, the individual $U(1)$s on each brane
conspire to form a non-abelian $U(N)$ \cite{Wittenbound,Douglas} thus
filling in the non-abelian gap in the supermultiplets. The reader is
referred to Polchinski's lectures \cite{TASI} for a full account of the
$D$-brane revolution.    

The bosonic sector of the  worldvolume actions for these $D$-branes is known
to be described by a Born-Infeld Lagrangian . From a bosonic point of
view, this can be deduced from their open-string interpretation
\cite{Leigh,Cederwall1}; from a fermionic point of view, this is suggested
by the solitonic interpretation: half of the spacetime supersymmetries are
non-linearly realized on the soliton worldvolume \cite{Luthree}. The
programme to display the covariant kappa symmetric actions for these Type
$II$ $p$-branes, spurred on by their interpretation as $D$-branes,  has
only recently yielded any results: the $D=10$ superthreebrane
\cite{Luthree} equations in a Type $IIB$ supergravity background was
constructed in \cite{Cederwall2}; all the $D=10$ $D$-branes in a flat
background in \cite{Aganagic} and all the $D=10$ $D$-branes in supergravity
backgrounds in \cite{Sundell,Kappa}. A different approach in
\cite{Howesezgin,Howesezginp} provides them in a version with both manifest
spacetime and worldvolume supersymmetry \cite{Howesezgin}, and also the
related $D=11$ superfivebrane \cite{Howesezginp}. The action of
\cite{Aganagic} is given by  \begin{equation}
S=-\int d^{p+1}\xi \sqrt{-det(G_{\mu\nu}+{\cal F}_{\mu\nu})}+ \int
\Omega_{p+1} 
\end{equation}
where
\begin{equation}
G_{\mu\nu} = \eta_{mn} \Pi^m{}_\mu \Pi^n{}_\nu,
\end{equation}
\begin{equation}
\Pi^m{}_\mu = \partial_\mu X^m - \bar\theta \Gamma^m \partial_\mu \theta,
\end{equation}
and
\begin{equation}
{\cal F}_{\mu\nu} = F_{\mu\nu} - b_{\mu\nu},
\end{equation}
where $F_{\mu\nu} = \partial_\mu A_\nu - \partial_\nu A_\mu$ and where
\begin{equation}
b = \bar\theta \Gamma_{11} \Gamma_m d\theta \left(d X^m - {1\over 2} \bar
\theta \Gamma^m d\theta\right).
\end{equation}
This is the formula for $p$ even.  When $p$ is odd, $\Gamma_{11}$ is replaced by
$\tau_3$. $\Omega_{p + 1}$ is a $(p + 1)$-form Wess--Zumino-type
term.  $S_1$ and $S_2$ are separately invariant under the global $IIA$ or
$IIB$ super-Poincare group as well as under $(p + 1)$-dimensional general
coordinate transformations.  However, local kappa symmetry is achieved
by a subtle conspiracy between them \cite{Aganagic}, just as in the case of
super $p$-branes with scalar supermultiplets (\ref{supermembrane}).

To date, no kappa symmetric actions have been written down for the $D=11$ (or
Type $IIA$) superfivebrane or the $D=10$ heterotic fivebrane. 

Finally, it has been suggested that there might be a $D=11$ superninebrane
\cite{Papa,Howesezgin} denoted by ``$?$'' on the brane scan of Table
\ref{scan}, which in turn suggests a formulation of $D=11$ supergravity
with an $11$-form field strength. This might then explain both the Type
$IIA$ supereightbrane and the Romans formulation of Type $IIA$
supergravity from an eleven-dimensional point of view. If so, one might
expect $R^6$ corrections to the nine brane Bianchi identity coming from
ninebrane worldvolume Lorentz anomalies just as one obtains $R^4$
corrections to the fivebrane Bianchi identities coming from fivebrane
worldvolume anomalies \cite{Duffliuminasian}. This, in turn would mean
$K_0 R^6$ corrections to the $D=11$ supergravity action (written as an
integral over a $12$-manifold) in addition to the $K_4 R^4$ already known
\cite{Duffliuminasian}, where $K_0$ and $K_4$ are respectively dual to the
ninebrane and fivebrane field strengths.  However, if one follows
the rule (\ref{scalars}) for counting transverse dimensions, such a ninebrane
would have to have one scalar on its worldvolume \cite{Howesezgin} and that
means the zero modes might have to include the ($n=1,d=10$) supergravity
multiplet itself!  Consequently, this would violate the usual folklore that
there are no spin $2$ zero modes on the worldvolume. (One could also add 
($n=1,d=10$) vector multiplets since these have no scalars.) On the other
hand, it might fit in nicely with the work of Horava and Witten
\cite{Horava} who have shown that the $E_8 \times E_8$ heterotic string
corresponds to the $D=11$ theory on $R^{10} \times S^1/Z_2$, so that the
space does indeed have ``ninebrane" boundaries.

\section{Aspects of quantization} 
\subsection{The
lightcone gauge: area-preserving diffeomorphisms} 
\label{lightcone}
In the string theory, the lightcone gauge is convenient for quantization
because it allows the elimination of all unphysical degrees of freedom and
unitarity is guaranteed.  Of course, one loses manifest Lorentz invariance
and one must be careful to check that it is not destroyed by quantization.

In membrane theory, however, the lightcone gauge does not eliminate all
unphysical degrees of freedom.  Let us split

\setcounter{equation}{0}
\begin{equation}
X^{\mu} = \biggl\{ X^{\pm} = \frac{1}{\sqrt
2}~\biggl(X^0\pm~X^{D-1}\biggr);~~~X^I, I = 1\ldots,(D-2)\biggr\}
\end{equation}
and then set
\begin{equation}
{\dot X}^+ = p^+
\end{equation}
where the dot denotes differentiation with respect to $\tau$, the time
coordinate on the worldvolume.  One can then solve for $X^-$ leaving the
$(D-1)$ variables $X^I$.  For membranes, however, only $(D-d)$ variables are
physical as explained in section (\ref{scan}).  Thus the lightcone gauge must
leave a residual gauge invariance \cite{Hoppe}.  Let us focus our attention
on  a $d = 3$ bosonic membrane in flat spacetime.  The lightcone action turns
out to be  
\begin{equation}
S= \frac{1}{2} \int d\tau \int d^2\sigma\biggl
[ \biggl( D_0 X^I\biggr)^2 - det
\partial_aX^I\partial_bX^I\biggr)
\end{equation}
where
\begin{equation}
D_0 = \partial_0 +u^a(\sigma,\tau)\partial_a;~~\partial_0
=\frac{\partial}{\partial\tau},~\partial_a =
\frac{\partial}{\partial\sigma^a}, a = 1,2~~.
\end{equation}
\noindent
$D_0$ is a ``covariant time derivative" with ``gauge field" $u^a$ satisfying
\begin{equation}
\partial_a u^a = 0~~.
\label{u}
\end{equation}
The action possesses a gauge invariance that ensures that only $(D - 3)$ of
the $(D -2)$ $X^I$ are physical.  For a membrane of spherical topology, the
solution to (\ref{u}) is 
\begin{equation}
u^a = \epsilon^{ab}\partial_b\omega~~
\label{solution}
\end{equation}
If we now introduce the Lie bracket
\begin{equation}
\{f,g\} = \epsilon^{ab}\partial_af~\partial_bg
\label{Lie}
\end{equation}
the action may be rewritten in polynomial form
\begin{equation}
S =
\frac{1}{2}\int d\tau\int d^2\sigma \biggl[\biggl(\partial_0X^I-\lbrace
\omega,X^I\rbrace\biggr)^2-\frac{1}{2}\lbrace X^I,X^J,\rbrace
\lbrace X^I,X^J\rbrace\biggr]~~.
\end{equation}
Remarkably, this looks like a $(D-1)$ dimensional Yang-Mills theory
dimensionally reduced to one time dimension.  To be more precise, consider
the correspondence
\begin{eqnarray}
X^I&\rightarrow&A^I\\
\omega&\rightarrow&A_o\\
\{,\}&\rightarrow&[,]\\
\int~d^2\sigma&\rightarrow& tr
\label{Yang}
\end{eqnarray}
then (\ref{Yang}) describes a Yang-Mills theory with infinite dimensional
gauge group.  Including the fermions, we get the supersymmetric action
\begin{equation} 
S = \frac{1}{2}\int d\tau tr\biggl\{ \biggl(D_0 A^I\biggr)^2
- \frac{1}{2}
\biggl[A^I,A^J\biggr]~\biggl[A^J,A^I\biggr]+i{\bar\lambda}D_0\lambda
+i{\bar\lambda} {\gamma}^I~\biggl[A^{I,}\lambda\biggr]\biggr\}
\label{action}
\end{equation} 
where the fields $(A_0,A^I)$ and $\lambda^{\alpha}$ are all in the adjoint
representation of $SU(\infty)$. This group is, in fact, the subgroup of
the worldvolume diffeomorphism group that preserves the Lie bracket
(\ref{Lie}) and is known as the group of area-preserving diffeomorphisms.
For spherical membranes, this group is given \cite{Hoppe} by lim $N
\rightarrow \infty$ $SU(N)$.  The toroidal case is discussed in
\cite{Floratos}.  Here, as for Riemann surfaces of genus $g \geq 1$, one must
take into account that the solution (\ref{solution}) is valid only locally.

One can generalize these results to the $d = 3$ supermembranes \cite{deWit1}
in $D = 4,5,7$ and $11$, and one finds super Yang-Mills quantum mechanical
models corresponding to the dimensional reduction of super Yang-Mills in $D
= 3,4,6$ and $10$, which provides yet another way of understanding the
allowed values of $ D$.  (One might conjecture a similar relationship
between the $d > 3$ membranes and quantum mechanical models, but this time
the gauge symmetry could not be of the Yang-Mills type.  It has been
suggested \cite{Bergshoeff2} that they are given by infinite dimensional
non-Abelian antisymmetric tensor gauge theories.)

\subsection{Massless states and anomalies}
\label{massless}

The connection between supersymmetric Yang-Mills quantum-mechanical models
and supermembranes has proved useful in the debate about the spectrum of the
supermembrane and the issue of whether or not it contains massless
particles.  The membrane supercharge and the Hamiltonian are given by
\[
Q=\int~d^2\sigma\biggl[ P_I\gamma^I+\frac{1}{2}~\lbrace X^I,
X^J\rbrace\gamma^{IJ}\biggr]
\]
\begin{equation}
H=\int~d^2\sigma~\biggl[\frac{1}{2}P_I^2+\frac{1}{2}
h^{ab}\lambda^{\dagger}{\gamma_{a}\gamma\partial_{b}}\lambda 
+ \frac{1}{4}~\biggl\{
X^I,X^J\biggr\}~\biggl\{ X^I,X^J\biggr\} \biggr]
\end{equation}
where
\begin{equation}
\gamma_a=\partial_aX^I\gamma^I
\end{equation}
and
\begin{equation}
\gamma =
\frac{i}{2}\epsilon^{ab}\partial_aX^I\partial_bX^J\gamma^{IJ}~~~.
\end{equation}
One sees, classically, that the membrane Hamiltonian is positive
semi-definite and that the zero energy configurations are those for which the
potential energy
\begin{equation}
h = det \partial_a
X^I \partial_b X^J=\frac{1}{2}~\biggl\{
X^I,X^J\biggr\}~\biggl\{ X^I,X^J\biggr\} 
\end{equation}
vanishes.  These are just the ``collapsed" membranes with zero area. 
(Intuitively, one would expect membranes and higher dimensional objects to be
different from strings in this respect, because of the ability to deform the
object without increasing its volume.)  Since the charges $Q$ and $H$ obey
\begin{equation}
\{Q^A,Q^{\dagger}_B\} = 2\delta^A_B H
\end{equation}
and the quantum Hamiltonian is positive semi-definite and the zero-energy
states are those annihilated by $Q$ i.e. the supersymmetric states, it seems
reasonable to conclude that the supermembrane contains massless states
provided that supersymmetry is not broken \cite{Inami}.

At the level of perturbation theory, Bars, Pope and Sezgin \cite{Bars1} have
shown that this expectation is borne out.  They conclude that in flat $D =
11$ spacetime, the massless sector of the supermembrane is just $D = 11$
supergravity, just as the massless sector of the superstring is $D = 10$
supergravity.  Curiously, they find that although the other super extended
objects in the $R$, $C$ and $H$ sequences also have massless particles, the
corresponding supermultiplet is {\it not} the supergravity multiplet. The
apparent mismatch between the background fields in the supermembrane action
and the massless states for the $R$, $C$ and $H$ sequences, suggests that
these theories are inconsistent. As discussed in section \ref{lightcone}, the
price to be paid for a unitary gauge is the loss of manifest Lorentz
invariance.  Sure enough, it has been shown by Bars \cite{Bars2} and Bars and
Pope \cite{Bars3} that all the members of the $R$, $C$ and $H$ sequences
suffer from anomalies\footnote{Note that this refers to the ``neutral''
$p$-branes \cite{Bergshoeff1} and not the ``heterotic'' $p$-branes.}. 

The major problem with the anomaly-free eleven-dimensional supermembrane,
however, as pointed out in \cite{deWit1,deWit2} is that it seems to have a
continuous rather than discrete spectrum as a consequence of the flat
directions of the potential discussed in section (\ref{lightcone}).
Banks, Fischler, Susskind and Shenker \cite{BanksM} have recently turned this
around to make a virtue of necessity, and suggested that $M$-theory should be
identified with the supersymmetric  $SU(\infty)$ matrix model desribed in
section (\ref{lightcone}). Following Townsend \cite{TownsendD}, they
identify the membrane with an infinite collection of $D$ $0$-branes since
the action for $N$ coincident $D$ $0$-branes is given precisely by the
above $SU(N)$ matrix models. In their picture, a continuous spectrum is to
be welcomed!

\section{P-Branes as Solitons}
\label{pbranes}

The study of supermembranes received further impetus with their
interpretation as {\it solitons}.  Solitons
\cite{Coleman,Jackiw,Rajaraman}, sometimes called {\it topological defects}
\cite{Vilenkin} are important in quantum field theory for a variety of
reasons. Their existence means that the full non-perturbative theory may
have a much richer structure than is apparent in perturbation theory. For
example, the electrically charged {\it elementary} particle spectrum may
need to be augmented by magnetically charged {\it solitonic} particles. The
former carry a Noether charge following from the equations of motion while
the latter carry a topological charge associated with the Bianchi
identities. That these magnetic monopoles are intrinsically
non-perturbative is apparent from their mass formula which depends
inversely on the coupling constant. The spectrum might also contain {\it
dyons}, particles carrying both electric and magnetic charge.

In four spacetime dimensions, extended solitonic objects, strings and domain
walls, are also possible. In modern parlance, they might be known as
{\it $p$-branes} with $p=0,1,2,$ where $d=p+1$ is the dimension of the 
worldvolume swept out by the soliton. In these lectures, we shall use the word
{\it solitons} to mean any such non-singular lumps of field energy which
solve the field equations, which have finite mass per unit $p$-volume and
which are prevented from dissipating by some topological conservation law.
Their existence, which in some grand unified theories is actually mandatory,
has far-reaching implications both at the microscopic and cosmic scales
\cite{Hindmarsh}.

However, it is now commonly believed that ordinary quantum field theory is
inadequate for a unification of all the forces including gravity and that it
must be supplanted by superstring theory or something even grander such as
$M$-theory
\cite{Howe,Luduality,Hulltownsend,Townsendeleven,Wittenvarious,%
Duffliuminasian,Becker1,Schwarzpower,Aharony,Schwarzreview,DuffM,BanksM}.
If this is true then soliton solutions of string theory and $M$-theory must
be even more important. Although the revival of the string idea is now
twelve years old, it is only recently that much attention has been devoted
to the subject of solitons \cite{Khuristring}. This interest has in part
been brought about by the realization that the really crucial questions of
string theory: ``How does the string choose a vacuum state?''; ``How does
the string break supersymmetry?''; ``How does the string cope with the
cosmological constant problem?'' cannot be answered within the framework of
a weak coupling perturbation expansion. 

As we have seen, the development of supermembranes did not at first proceed
in the soliton direction, despite the soliton approach of the original paper
\cite{Hughes}. The next soliton development came when Townsend
\cite{Townsend4} pointed out that not merely the $D=6$ threebrane but all the
points on the ${\cal H,C,R}$ sequences correspond to topological defects of
some globally supersymmetric  field theory which break half the spacetime
supersymmetries. This partial breaking of supersymmetry, already discussed in
\cite{Hughes}, is a key idea and is intimately connected with the worldvolume
kappa symmetry which allows one to gauge away half of the fermionic degrees
of freedom. He conjectured that the $p$-branes in the ${\cal O}$ sequence
would also admit such a solitonic interpretation within the context of
supergravity. 

The first hint in this direction came from  Dabholkar et al. \cite{Dabholkar} 
who presented a multi-string solution which in $D=10$ indeed breaks half the
supersymmetries. They obtained the solution by solving the low-energy $3$-form
supergravity equations of motion coupled to a string $\sigma$-model source and
demonstrated that it saturated a Bogomol'nyi bound and satisfied an associated
zero-force condition, these properties being intimately connected with the
existence of unbroken spacetime supersymmetry.  We shall shortly rederive
this string solution and point out the existence of the Bogomol'nyi bound 
between the ADM mass per unit length and charge per unit length and discuss
the zero-force condition which arises from the preservation of half the
spacetime supersymmetries.

However, this $D=10$ string was clearly not the soliton anticipated by
Townsend because it described a singular configuration with a $\delta$-function
source at the string location. Moreover, its charge per unit length
$e_2$ was an ``electric'' Noether charge associated with the equation of
motion of the antisymmetric tensor field rather than a ``magnetic''
topological charge associated with the Bianchi identities. Consequently, in the
current literature on the subject, this solution is now referred to as
the ``fundamental'' or ``elementary'' string. Similarly, the
supermembrane solution of $D=11$ supergravity found in \cite{Duffstelle} did not
seem to be solitonic either because it was also obtained by coupling to a
membrane $\sigma$-model source \footnote{Curiously, however, the curvature 
computed from its $\sigma$-model metric is finite at the location of the
source, in contrast to the case of the elementary string
\cite{Gibbonsdufftownsend}.}.

The next major breakthrough for $p$-branes as solitons came with the paper of
Strominger \cite{Strominger1}, who showed that $D=10$ supergravity coupled to
super Yang-Mills (without a $\sigma$-model source), which is the field theory
limit of the heterotic string, admits as a solution the heterotic
fivebrane. In contrast to the elementary string, this fivebrane is a
genuine soliton, being everywhere nonsingular and carrying a topological
magnetic charge $g_6$. A crucial part of the construction was a Yang-Mills
instanton in the four directions transverse to the fivebrane. He went on to
suggest a complete strong/weak coupling duality with the strongly coupled
string corresponding to the weakly coupled fivebrane and vice-versa, thus
providing a solitonic interpretation of the string/fivebrane duality
conjecture. In this form, string/fivebrane duality is in a certain sense an
analog of the Montonen-Olive conjecture discussed in section
(\ref{SandT}), according to which the magnetic monopole states of
four-dimensional spontaeously broken supersymmetric Yang-Mills theories
may be viewed from a dual perspective as fundamental in their own right
and in which the roles of the elementary and solitonic states are
interchanged.

This strong/weak coupling idea received further support from the
point of view of Poincare duality in \cite{Luremarks}. There it was shown
that the just as the string loop expansion parameter is given by ${\rm
g}_2=e^{\phi_0}$, where $\phi_0$ is the dilaton vev, so the analogous 
fivebrane parameter is given by ${\rm g}_6=e^{-\phi_0/3}$ and hence that
${\rm g}_6={\rm g}_2^{-1/3}$. The same paper also established a Dirac
quantization rule $\kappa^2 T_2T_6=n\pi$, where $n$ is an integer, relating the
fivebrane tension $T_6$ to the string tension $T_2$, which followed from the
corresponding rule for the electric and magnetic charges generalized to extended
objects \cite{Nepomechie,Teitelboim} $e_2 g_6=2n\pi$.

In keeping with the viewpoint that the fivebrane may be
regarded as fundamental in its own right, the elementary
fivebrane solution \cite{Luelem} was constructed  by coupling the $7$-form version of supergravity to a
fivebrane $\sigma$-model source in analogy with the elementary string. This
carries an electric charge $\tilde e_6$. String/fivebrane duality then
suggested that by coupling the $7$-form version of supergravity to super
Yang-Mills (without a $\sigma$-model source), one ought to find a
nonsingular heterotic string soliton carrying a topological magnetic charge
$\tilde g_2$. Here one would expect an eight-dimensional Yang-Mills
instanton in the eight directions transverse to the string. This was indeed
the case \cite{Lustrings}, but scaling arguments required an unconventional
Yang-Mills Lagrangian, quartic in the field strengths, which, however, is
only to be expected in a fivebrane loop expansion \cite{Lustrings,Luloop}.

Somewhat surprisingly, the elementary fivebrane, as pointed out
by Callan, Harvey and Strominger \cite{Callan1,Callan2}, could
also be regarded as a soliton when viewed from the dual perspective,
with $g_6=\tilde e_6$. In other words, it provides a nonsingular
solution of the source-free $3$-form equations even without the presence of
Yang-Mills fields. By the same token, when viewed from the dual perspective,
the elementary string provides a nonsingular solution of the source-free
$7$-form equationswith $\tilde g_2=e_2$ \cite{Luremarks,Luelem}. Then Callan
Harvey and Strominger showed that similar results also appear in both Type
$IIA$ and Type $IIB$ string theories \cite{Callan1,Callan2}; they also admit
fivebrane solitons.

\subsection{The elementary string}
\label{elementary}

We begin by recalling the elementary string solution of \cite{Dabholkar}.
We want to find a vacuum-like supersymmetric configuration with $D = 2$
super-Poincare symmetry from the 3-form version of $D = 10, N = 1$ supergravity
theory. As usual, the fermionic fields should vanish for this configuration.
We start by making an ansatz for the $D = 10$ metric $g_{MN}$, 2-form
$B_{MN}$ and dilaton $\phi$ ($M = 0, 1, \cdots,9$) corresponding to the
most general eight-two split invariant under $P_2 \times SO(8)$, where $P_2$ is
the $D = 2$ Poincare group. We split the indices
\begin{equation}
{x^M = (x^\mu, y^m),}
\label{split}
\end{equation}
where $\mu = 0,1$ and $m = 2, \cdots, 9$, and write the line element as
\begin{equation}
{ds^2 = e^{2A} \eta_{\mu\nu} dx^\mu dx^\nu + e^{2B}\delta_{mn} dy^mdy^n ,}
\label{element}
\end{equation}
and the two-form gauge field as
\begin{equation}
{B_{01} = - e^C .}
\end{equation}
All other components of $B_{MN}$ and all components of the gravitino $\psi_M$ and
dilatino $\lambda$ are set zero. $P_2$ invariance requires that the arbitrary
functions $A, B$ and $C$ depend only on $y^m$; $SO(8)$ invariance then requires
that this dependence be only through $y = \sqrt {\delta_{mn} y^m y^n}$.
Similarly, our ansatz for the dilaton is
\begin{equation}
{\phi = \phi(y).}
\end{equation}
As we shall now show, the four arbitrary functions $A, B, C,$ and $\phi$ are
reduced to one by the requirement that the above field configurations preserve
some unbroken supersymmetry. In other words, there must exist Killing spinors
$\varepsilon$ satisfying \cite{Dabholkar} 
\begin{equation}
{\delta \psi_M = D_M
\varepsilon + {1\over 96}\,e^{-\phi/
2}\big(\Gamma_M\,^{NPQ} - 9~\delta_M\,^N \Gamma^{PQ}\big)
 H_{NPQ}~
\varepsilon = 0,}
\end{equation}
\begin{equation}
{\delta \lambda = - {1\over2\sqrt{2}}~\Gamma^M
\partial_M \phi
\varepsilon + {1\over 24\sqrt{2}}~e^{-\phi/2}~\Gamma^{MNP}
H_{MNP}~\varepsilon = 0,}
\end{equation}
where
\begin{equation}
{H_{MNP} = 3\partial_{\lbrack M} A_{NP\rbrack}.}
\end{equation}
Here $\Gamma_A$ are the $D = 10$ Dirac matrices satisfying
\begin{equation}
{\lbrace \Gamma_A, \Gamma_B \rbrace = 2\eta_{AB}.}
\end{equation}
$A, B$ refer to the $D = 10$ tangent space, $\eta_{AB} = (-, +,\cdots, +)$, and
\begin{equation}
{\Gamma_{AB\cdots C} = \Gamma_{\lbrack A}
\Gamma_{B \cdots}
\Gamma_{C\rbrack},}
\end{equation}
thus $\Gamma_{AB} = {1\over 2}~(\Gamma_A \Gamma_B - \Gamma_B\Gamma_A)$, etc.
The $\Gamma$'s with world-indices $P, Q, R, \cdots$  have been converted using
vielbeins $e_M\,^A$. We make an eight-two split
\begin{equation}
{\Gamma_A = (\gamma_\alpha\otimes 1, \gamma_3\otimes \Sigma_a),}
\end{equation}
where $\gamma_\alpha$ and $\Sigma_a$ are the $D = 2$ and $D = 8$ Dirac
matrices, respectively. We also define 
\begin{equation}
{\gamma_3 = \gamma_0\gamma_1,} 
\end{equation}
so that $\gamma_3^2 = 1$ and 
\begin{equation}
{\Gamma_9 = \Sigma_2 \Sigma_3 \cdots \Sigma_9 ,}
\end{equation}
so that $\Gamma_9^2 = 1$. The most general spinor consistent
with $P_2 \times SO(8)$ invariance takes the form
\begin{equation}
{\varepsilon (x, y) = \epsilon \otimes \eta ,}
\end{equation}
where $\epsilon$ is a spinor of $SO(1,1)$ which may be further decomposed into
chiral eigenstates via the projection operators $(1\pm \gamma_3)$ and $\eta$ is
an $SO(8)$ spinor which may  further be decomposed into chiral eigenstates via
the projection operators $(1\pm\Gamma_9)$. The $N=1, D=10$ supersymmetry
parameter is, however, subject to the ten-dimensional chirality
condition
\begin{equation}
{\Gamma_{11}~\varepsilon = \varepsilon ,}
\end{equation}
where $\Gamma_{11} = \gamma_3\otimes\Gamma_9$ and so the $D=2$ and $D=8$
chiralities are correlated.

Substituting the ansatz into the supersymmetry transformation rules
leads to the solution \cite{Dabholkar}
\begin{equation}
{\varepsilon = e^{3\phi/8}\epsilon_0\otimes\eta_0,}
\end{equation}
 where $\epsilon_0$ and $\eta_0$ are constant spinors satisfying
\begin{equation} 
{(1 -\gamma_3)\epsilon_0 =0,\quad (1 - \Gamma_9) \eta_0 = 0,}
\label{chiral}
\end{equation}
and where
\begin{equation}
A= {3\phi\over 4} +c_A ,
\end{equation}
\begin{equation}
B= -{\phi\over 4} + c_B ,
\end{equation}
\begin{equation}
C= 2\phi +2 c_A ,
\end{equation}
where $c_A$ and $c_B$ are constants. If we insist that the
metric is asymptotically Minkowskian, then
\begin{equation}
{c_A = -~{3\phi_0\over 4},\quad c_B = ~{\phi_0\over 4},}
\end{equation}
where $\phi_0$ is the value of $\phi$ at infinity {\it i.e.} the dilaton
vev $\phi_0 =~<\phi>$. The condition (\ref{chiral}) means that one half of
the supersymmetries are broken.

At this stage the four unknown functions $A$, $B$, $C$ and
$\phi$ have been reduced to one by supersymmetry. To determine $\phi$, we must
substitute the ansatz into the field equations which follow from the action
$I_{10}({\rm string}) + S_2$ where $I_{10}({\rm string})$ is the bosonic sector
of the 3-form version of $D =10, N = 1$ supergravity given by
\begin{equation}
{I_{10}({\rm string}) = {1\over 2\kappa^2}
\int d^{10}x~\sqrt{-g}~\bigg(R - {1\over 2}
(\partial\phi)^2 - {1\over 2\cdot 3!}~e^{-\phi} H^2\bigg),}
\end{equation} 
and $S_2$ is the string $\sigma$-model action. Up until now we have employed
the canonical choice of metric for which the gravitational action is the
conventional Einstein-Hilbert action. This metric is related to the metric
appearing naturally in the string $\sigma$-model by 
\begin{equation}
{g_{MN}({\rm string}~\sigma{\rm\!-\!model}) =
e^{\phi/2} g_{MN}({\rm canonical}),}
\end{equation}
In canonical variables, therefore, the string $\sigma$-model action is
given by
\begin{eqnarray}
S_2 &=& - T_2 \int d^2 \xi \bigg({1\over 2}
\sqrt{-\gamma}~
\gamma^{ij}\partial_i X^M \partial_j X^N g_{MN}~
e^{\phi/2} - 2~\sqrt{-\gamma}
\nonumber \\
&& ~~~~~~~~~~~
+{1\over 2!}~\varepsilon^{ij} \partial_i X^M \partial_j X^N
 B_{MN}\bigg).
\end{eqnarray}
We have denoted the string tension by $T_2$. The supergravity field equations
are
\begin{eqnarray}
R^{MN} \!\!\!\! &-&\!\!\!\!
{1\over 2}~\bigg(\partial^M
\phi~\partial^N \phi - {1\over 2}~g^{MN} (\partial \phi)^2 \bigg)
- {1\over 2}~g^{MN} R
\nonumber \\
 \!\!\!\!&-&\!\!\!\! {1\over 2\cdot 2!} \bigg(H^M\,_{PQ} H^{NPQ} - {1\over6}
{}~g^{MN}
H^2\bigg) e^{-\phi}
=\kappa^2 T^{MN}(\rm string),
\end{eqnarray}
where
\begin{equation}
T^{MN}({\rm string}) = -T_2 \int d^2\xi\sqrt{-\gamma}
{}~\gamma^{ij}
\partial_i X^M \partial_j X^N e^{\phi/2}~{\delta^{10} (x - X)
\over \sqrt{-g}},
\end{equation}
\begin{equation}
\partial_M (\sqrt{-g}~e^{-\phi} H^{MNP}) =2\kappa^2 T_2 \int d^2
\xi~\varepsilon^{ij} \partial_i X^N \partial_j X^P
\delta^{10} (x - X),
\end{equation}
\begin{eqnarray}
 &&
\partial_M (\sqrt{-g}g^{MN}\partial_N\phi) + {1\over 2\cdot
3!}~e^{-\phi} H^2 =
\nonumber \\
&& ~~~~~~~~~~~= {\kappa^2 T_2\over 2}
\int d^6 \xi \sqrt{-\gamma}~\gamma^{ij}\partial_i X^M
\partial_j X^N g_{MN}
e^{\phi/2}{\delta^{10} (x - X)}.
\end{eqnarray}
Furthermore, the string field equations are
\[
\partial_i (\sqrt{-\gamma}~\gamma^{ij}
\partial_j X^N
g_{MN}~e^{\phi/2} ) - {1\over 2}~\sqrt {-\gamma}~\gamma^{ij}
\partial_i X^N
\partial_j X^P \partial_M (g_{NP}~e^{\phi/2} )-
\]
\begin{equation}
{1\over 2}~\varepsilon^{ij}
\partial_i X^N \partial_j X^P H_{MNP} = 0, 
\end{equation}
and
\begin{equation}
\gamma_{ij} = \partial_i X^M \partial_j X^N g_{MN}
e^{\phi/2}.
\end{equation}
To solve these coupled supergravity-string equations we make
the static
gauge choice
\begin{equation}
{X^\mu = \xi^\mu, \quad \mu = 0,1}
\end{equation}
and the ansatz
\begin{equation}
{X^m=Y^m = {\rm constant}, \qquad m=2,...,9.}
\end{equation}
As an example, let us now substitute the ansatz into and
the 2-form equation. We find
\begin{equation}
{\delta^{mn} \partial_m \partial_n e^{-2\phi} =
 - 2\kappa^2 T_2
e^{-\phi_0/2}\delta^8 (y),}
\end{equation}
and hence
\begin{equation}
{e^{-2\phi} = e^{-2\phi_0} \left(1 + {k_2\over y^6}
\right),}
\end{equation}
where the constant $k_2$ is given by
\begin{equation}{k_2 \equiv {\kappa^2 T_2\over 3\Omega_7}
{}~e^{3\phi_0/2},}
\label{single}
\end{equation}
and $\Omega_n$ is the volume of the unit $n$-sphere $S^n$. One
 may verify by using the expressions for the
Ricci tensor $R^{MN}$ and Ricci scalar $R$ in terms of $A$ and $B$
\cite{Khuristring} that all the field equations are reduced to a
 single equation (\ref{single}).

Having established that the supergravity configuration preserves half the
supersymmetries, we must also verify that the string configuration also
preserve these supersymmetries. As discussed in \cite{Bergshoeffgamma},
the criterion is that in addition to the existence of Killing spinors
 we must also have 

\begin{equation} {(1 - \Gamma)\varepsilon = 0,}
\end{equation} where the choice of sign is dictated by the choice of the
sign in the Wess-Zumino term in $S_2$, and where 
\begin{equation}
{\Gamma \equiv {1\over 2! \sqrt{-\gamma}}
{}~\varepsilon^{ij}
\partial_i X^M \partial_j X^N \Gamma_{MN}.}
\label{gamma}
\end{equation}
Since $\Gamma^2 = 1$ and tr $\Gamma = 0,\, {1\over 2} (1 \pm
\Gamma)$ act as projection operators. For our solution, we find  that
\begin{equation}
{\Gamma = \gamma_3 \otimes 1,}
\end{equation}
and hence (\ref{gamma}) is satisfied.  This explains, from a string point of
view, why the solutions we have found preserve just half
the supersymmetries. It originates from the fermionic kappa symmetry of
the superstring action. The fermionic zero-modes on the worldvolume are
just the Goldstone fermions associated with the broken supersymmetry.

As shown in \cite{Dabholkar}, the elementary string solution saturates a
Bogolmol'nyi bound for the mass per unit length
\begin{equation}
{{\cal M}_2 = \int d^8 y~\theta_{00},}
\end{equation}
where $\theta_{MN}$ is the total energy-momentum pseudotensor of the combined
gravity-matter system. One finds 
\begin{equation}
{\kappa {\cal M}_2 \geq {1\over\sqrt{2}} |e_2| e^{ \phi_0/2},}
\end{equation}
where $e_2$ is the Noether ``electric '' charge whose conservation follows the
equation of motion of the 2-form, namely
\begin{equation}
{e_2 = {1\over {\sqrt{2}\kappa}} \int\limits_{S^7}e^{-\phi}\,^{\ast} H,}
\end{equation} 
where $^{\ast}$ denotes the Hodge dual using the canonical metric
and the integral is over an asymptotic seven-sphere surrounding the string. We
find for our solution that
\begin{equation}
{{\cal M}_2 = e^{\phi_0/2}~T_2,}
\label{mass}
\end{equation}
and
\begin{equation}
{e_2 = \sqrt{2}\kappa~T_2.}
\label{charge}
\end{equation}
Hence the bound is saturated. This provides another way, in addition to
unbroken supersymmetry, to understand the stability of the solution.

\subsection{The solitonic fivebrane}
\label{soliton}
The elementary string discussed above is a solution of the
coupled field-string system with action $I_{10}({\rm string})+ S_2$.  
As such it exhibits $\delta$-function singularities at $y = 0$.  It is
characterized by a non-vanishing Noether electric charge $e_2$.  By
contrast, we now wish to find a solitonic fivebrane, corresponding to a
solution of the source free equations resulting from $I_{10}({\rm string})$
alone and which will be characterized by a non-vanishing topological
``magnetic'' charge $g_6$.

To this end, we now make an ansatz invariant under $P_6 \times SO(4)$. 
Hence we write (\ref{split}) and (\ref{element}) as before where now $\mu =
0, 1 \ldots 5$ and $m = 6, 7, 8, 9$.  The ansatz for the antisymmetric
tensor, however, will now be made on the field strength rather than on the
potential.  From section (\ref{elementary}) we recall that a non-vanishing
electric charge corresponds to
\begin{equation}
{{1\over \sqrt{2} \kappa} e^{-\phi}{}^{\ast}H
 = e_2 \varepsilon_7/\Omega_7,}
\end{equation}
where $\varepsilon_n$ is the volume form on $S^n$. Accordingly,
to obtain a non-vanishing magnetic charge, we make the ansatz
\begin{equation}
{{1\over \sqrt{2} \kappa} H = g_6
\varepsilon_3/\Omega_3.}
\end{equation}
 Since this is an harmonic form, $H$ can no longer be written globally as
the curl of $B$, but it satisfies the Bianchi identity.  It is now not
difficult to show that all the field equations are
satisfied. The solution is given by
\begin{equation}
e^{2\phi}=e^{2\phi_0}\left(1+{k_6\over y^2}\right),
\end{equation}
\begin{equation}
ds^2=e^{-(\phi-\phi_0)/2}\eta_{\mu\nu}dx^\mu dx^\nu+e^{3(\phi-\phi_0)/2}
\delta_{mn}dy^mdy^n,
\end{equation}
\begin{equation}
 H=2k_6 e^{\phi_0/2}\varepsilon_3,
\end{equation}
where $\mu,\nu=0,1,...,5$, $m,n=6,7,8,9$ and where
\begin{equation}{k_6={\kappa g_6 \over \sqrt{2}\Omega_3}e^{-\phi_0/2}.}
\end{equation}
It follows that the mass per unit 5-volume now saturates a bound
involving the magnetic charge
\begin{equation}{{\cal M}_6={1\over \sqrt{2}} \mid g_6 \mid
e^{-\phi_0/2}.}\end{equation}
Note that the $\phi_0$ dependence is such that ${\cal M}_6$ is
large
for small ${\cal M}_2$ and vice-versa.

The electric charge of the elementary solution and the magnetic charge of
the soliton solution obey a Dirac quantization rule
\cite{Nepomechie,Teitelboim} 
\begin{equation}{e_2 g_6 = 2 \pi n, \qquad n =
{\rm integer},} \end{equation}
and hence 
\begin{equation}{g_6 = 2\pi n/\sqrt{2}\kappa T_2.}
\end{equation}

\subsection{Type II solutions, $D$-branes and black branes}
\label{IIsolutions}

Let us begin in $D = 10$ with Type $IIA$ supergravity, whose bosonic action
is given by
\begin{eqnarray}
I_{10} (IIA)&=&{1\over 2\kappa^2}~\int d^{10} x \sqrt{-g}
\Bigg[R - {1\over 2}~(\partial\phi)^2 - {1\over 2.3!}~e^{-\phi} F_3\,^2
-{1\over 2.2!}~e^{3 \phi/2} F_2\,^2 
\nonumber \\
&& ~~~~~~~~~~~- {1\over 2.4!}~e^{\phi/2}
F'_4\,^2\Bigg]
- {1\over 4\kappa^2} \int F_4 \wedge F_4 \wedge A_2 ,
\end{eqnarray}
where
\begin{equation}
F'_4 = dA_3 + A_1 \wedge F_3 .
\end{equation}
Both the elementary string ($d = 2$) and fivebrane $(d = 6)$ solutions of $N =
1$ supergravity described above continue to provide solutions to Type IIA
supergravity, as may be seen by setting $F_2 = F_4 = 0$
\cite{Dabholkar,Callan1,Callan2}.  [This observation is not as obvious as it
may seem in the case of the elementary fivebranes or solitonic strings,
however, since it assumes that one may dualize $F_3$.  Now the Type IIA
action follows by dimensional reduction from the action of $D = 11$
supergravity which contains $F_4$.  There exists no dual of this action in
which $F_4$ is replaced by $F_7$ essentially because $A_3$ appears explicitly
in the Chern-Simons term $F_4 \wedge F_4 \wedge A_3$ \cite{Nicolaidual}. 
Since $F_4$~and~$F_3$ in $D = 10$ originate from $F_4$ in $D = 11$, this
means that we cannot {\it simultaneously} dualize $F_3$~and~$F_4$ but one may
do either {\it separately}.  By partial integration one may choose to have no
explicit $A_3$ dependence in the Chern-Simons term or no explicity
$A_2$ dependence, but not both.]  Furthermore, by setting $F_2 = F_3 = 0$ we
find elementary membrane $(d = 3)$ and solitonic fourbrane $(\tilde{d} = 5)$
solutions, and then by dualizing $F_4$, elementary fourbrane $(d = 5)$ and
solitonic membrane $(\tilde{d} = 3)$ solutions. Finally, by setting $F_3 =
F_4 = 0$, we find elementary particle $(d = 1)$ and solitonic sixbrane
$(\tilde{d} = 7)$ solutions and then by dualizing $F_2$, elementary sixbrane
$(d = 7)$ and solitonic particle $(\tilde{d} = 1)$ solutions
\cite{Luscan}.

Next we consider Type $IIB$ supergravity in $D = 10$ whose bosonic sector
consists of the graviton $g_{MN}$, a complex scalar $\phi$, a complex 2-form
$A_2$ (i.e with $d = 2$ or, by duality $d = 6$) and a real $4$-form $A_4$
(i.e with $d = 4$ which in $D = 10$ is self-dual).  Because of this
self-duality of the $5$-form field strength $F_5$, there exists no covariant
action principle.  Nevertheless we can apply the same logic to the
equations of motion and we find that the solution again falls into the
generic category.  First of all, by truncation it is easy to see that the
same string $(d = 2)$ and fivebrane $(d = 6)$ solutions of $N = 1$
supergravity continue to solve the field equations of Type IIB
\cite{Callan1,Callan2}.  On the other hand, if we set to zero $F_3$ and
solve the self-duality condition $F_5 = - ^{\ast}\!F_5$ then we find the
self-dual superthreebrane \cite{Luthree} discussed in more detail below.

All of the above elementary and solitonic solutions satisfy the
mass = charge condition. Our next task is to check for
supersymmetry. We begin by making the same ansatz as in sections
(\ref{elementary}) and (\ref{soliton}),  but this time
substitute into the supersymmetry transformation rules rather than the field
equations, and demand unbroken supersymmetry.  This reduces the four unknown
functions $A$, $B$, $C$ and $\phi$ to one.  We then  compare the results with
the known solutions.

For Type IIA supergravity with vanishing fermion background,
the gravitino transformation rule is
\[
\delta\psi_M=D_M \varepsilon +
{1\over 64}~e^{3\phi/4}
(\Gamma_M\,^{M_1M_2} - 14 \delta_M\,^{M_1} \Gamma^{M_2})
\Gamma^{11}
\varepsilon F_{M_1M_2}
\]
\[
 +{1\over 96}~e^{-\phi/2} (\Gamma_M\,^{M_1M_2M_3} - 9
\delta_M\,^{M_1}
\Gamma^{M_2M_3}) \Gamma^{11} \varepsilon F_{M_1M_2M_3}
\]
\begin{equation}
 +{i\over 256}~e^{\phi/4} (\Gamma_M\,^{M_1M_2M_3M_4} -
{20\over
3}~\delta_M\,^{M_1} \Gamma^{M_2M_3M_4}) \varepsilon
F_{M_1M_2M_3M_4}
\end{equation}
and the dilatino rule is
\[
\delta\lambda={1\over 4}~\sqrt{2}~D_M
\phi \Gamma^M \Gamma^{11}
\varepsilon + {3\over 16}~{1\over \sqrt{2}}~e^{3\phi/4} \Gamma^{M_1M_2}
\varepsilon F_{M_1M_2}
\]
\begin{equation}
 + {1\over 24}~{i\over \sqrt{2}}~e^{-\phi/2}
\Gamma^{M_1M_2M_3} \varepsilon F_{M_1M_2M_3}
 - {1\over 192}~{i\over \sqrt{2}}~e^{\phi/4}
\Gamma^{M_1M_2M_3M_4} \varepsilon F_{M_1M_2M_3M_4}
\end{equation}
where $\Gamma^M$ are the $D = 10$ Dirac matrices, where the covariant
derivative is given by
\begin{equation}
D_M = \partial_M + {1\over 4}~\omega_{MAB} \Gamma^{AB}
\end{equation}
with $\omega_{MAB}$ the Lorentz spin connection, where
\begin{equation}
\Gamma^{M_1M_2\ldots M_n} = \Gamma^{[M_1} \Gamma^{M_2} \ldots
\Gamma^{M_n]}
\end{equation}
and where
\begin{equation}
\Gamma^{11} = i \Gamma^0 \Gamma^1 \ldots \Gamma^9.
\end{equation}
Similarly the Type $IIB$ rules are
\begin{eqnarray}
&&\delta \psi_M = D_M \varepsilon+{i\over 4 \times
480}~\Gamma^{M_1M_2M_3M_4} \Gamma_M \varepsilon F_{M_1M_2M_3M_4}
\nonumber \\
&& ~~~~~~~~~~
+{1\over 96}~(\Gamma_M\,^{M_1M_2M_3} - 9 \delta_M\,^{M_1} \Gamma^{M_2M_3})
\varepsilon^{\ast} F_{M_1M_2M_3}
\end{eqnarray}
and
\begin{equation}
\delta\lambda = i \Gamma^M \varepsilon^{\ast} P_M - {1\over 24}~i
\Gamma^{M_1M_2M_3} \varepsilon F_{M_1M_2M_3}
\end{equation}
where
\begin{equation}
{P_M = \partial_M \phi/(1 - \phi^{\ast} \phi).}
\end{equation}
In the Type $IIB$ case, $\varepsilon$ is chiral
\begin{equation}
{\Gamma_{11} \varepsilon = \varepsilon.}
\end{equation}
The requirement of unbroken supersymmetry is that there exist Killing 
spinors $\varepsilon$ for which both $\delta\psi_M$~and~$\delta\lambda$
vanish. Substituting our ansatze into the transformation rules we find
that for every $1 \leq d \leq 7$ there exist field configurations which
break exactly half the supersymmetries.  This is just what one expects for
supersymmetric extended object solutions  and is intimately related to the
kappa symmetry discussed in section (\ref{type}) and the Bogomoln'yi bounds.
The important observation \cite{Luscan} is that the values of $A$, $B$ and
$\phi$ in terms of $C$ required by supersymmetry also solve the field
equations.  Thus in addition to the $D = 10$ super $(d - 1)$ branes already
known to exist for $d = 2$ (Heterotic, Type $IIA$ and Type $IIB$), $d = 4$
(Type $IIB$ only) and $d = 6$ (Heterotic,Type $IIA$ and Type $IIB$), we 
establish the existence of a Type $IIA$ superparticle $(d = 1)$, a Type
IIA supermembrane $(d = 3)$, a Type $IIA$ superfourbrane $(d = 5)$ and a
Type IIA supersixbrane $(d = 7)$ \cite{Luscan}.

It is perhaps worth saying a few more words about the
self-dual superthreebrane. By virtue of the (anti) self-duality condition
 $F_5=-^\ast F_5$, the electric Noether charge coincides with the topological
magnetic charge 
\begin{equation}
{e_4 = -g_4 .}
\end{equation}
(Note that such a condition is possible only in theories allowing a real
self-duality condition i.e. in 2 mod 4 dimensions, assuming Minkowski
signature. The self-dual string of \cite{Lublack} is another example.) 

Finally we count bosonic and fermionic zero modes. We know that one half of
the supersymmetries are broken, hence we have 16 fermionic zero modes. 
Regrouping these 16 fermionic zero modes, we get four Majorana spinors in $d
= 4$. Hence the $d = 4$ worldvolume supersymmetry is $N = 4$. Worldvolume
supersymmetry implies that the number of fermionic and bosonic on-shell
degrees of freedom must be equal, so we need a total of eight bosonic zero
modes. There are the usual six bosonic translation zero modes, but we are
still short of two. The two extra zero modes come from the excitation of the
complex antisymmetric field strength $G_{MNP}$. The equation of motion for
small fluctuations of the two-form potential $b$ in the soliton background 
is
\begin{equation}
D^P G_{MNP} = - {i\over 6}~F_{MNPQR} G^{PQR} .
\end{equation}
This is solved by
\begin{equation}
{b = e^{ik \cdot x} E \wedge de^{2A} ,}
\end{equation}
\begin{equation}
{G = db + i \ast (db) ,}
\end{equation}
where $k$ is a null vector in the two Lorentzian dimensions tangent to the
worldvolume. $E$ is a constant polarization vector orthogonal to $k$ but
tangent to the worldvolume and $\ast$ the Hodge dual in the worldvolume
directions. Although $G$ is a complex tensor, the zero-modes solution
gives only one real vector field on the worldvolume which provides the
other two zero modes. These two zero modes turn out to be pure
gauge at zero worldvolume momentum. Together with the other zero modes,
these fields make up the $d = 4, N = 4$ matter supermultiplet
$(A_{\mu}, \lambda^I,\phi^{[IJ]})$.

For all these solutions, the mass per unit $p$-volume was given by the
charge, as a consequence of the preservation half of the spacetime
supersymmetry.  However, it was recognized \cite{Luscan} that they were in
fact just the extremal mass=charge limit of more general non-supersymmetric
solutions found previously \cite{Horowitz1}. These
solutions, whose mass was greater than their charge, exhibit {\it event
horizons}: surfaces from which nothing, not even light, can escape. They
were {\it black} branes!  

Thus another by-product of these developments has been an
appreciation of the role played by black holes in particle physics and
string theory. In fact they can be regarded as black branes wrapped
around the compactified dimensions
\cite{Khurinew,Duffrahmfeld,Dufflupopeblack}. These black holes are tiny
$(10^{-35}$ meters) objects; not the multi-million solar mass objects that
are gobbling up galaxies. However, the same physics applies to both and
there are strong hints that $M$-theory may even clear up many of the
apparent paradoxes of quantum black holes \cite{BanksM} raised by Hawking.

One of the biggest unsolved mysteries in
string theory is why there seem to be billions of different ways of
compactifying the string from ten dimensions to four and hence
billions of competing predictions of the real world (which is like having no
predictions at all). Remarkably, Greene,  Morrison and Strominger
\cite{Greene} have shown that these wrappped around black branes actually
connect one Calabi-Yau vacuum to another. This holds promise of a dynamical
mechanism that would explain why the world is as it is, in other words, why
we live in one particular vacuum.

Furthermore, the $D$-brane technology has opened up a whole new chapter in the
history of supermembranes.  In particular, it has enabled Strominger and Vafa
\cite{Stromingervafa} to make a comparison of the black hole entropy
calculated from the degeneracy of wrapped-around black brane states with the
Bekenstein-Hawking entropy of an extreme black hole. Their agreement
provided the first microscopic explanation of black hole entropy. Moreover,
as Townsend \cite{Townsendeleven} had shown earlier, the extreme black hole
solutions of the ten-dimensional Type $IIA$ string (in other words, the
Dirichlet $0$-branes) were just the Kaluza-Klein particles associated with
wrapping the eleven-dimensional membrane around a circle. Moreover,
four-dimensional black holes also admit the interpretation of intersecting
membranes and fivebranes in eleven-dimensions
\cite{Strominger3,Townsendpapa,Klebanov2,Behrndt2,Gauntlett,Larsen}.  All
this holds promise of a deeper understanding of black hole physics via
supermembranes.

\subsection{Membranes and fivebranes in $D=11$}
\label{mandf}

In \cite{Duffstelle}, the supermembrane was recovered as an elementary
solution of  $D=11$ supergravity which preserves half of the spacetime
supersymmetry . Making the three/eight split $X^M=(x^{\mu},y^m)$ where
$\mu=0,1,2$ and $m=3,...,10$, the metric is given by 
\be
ds^2=(1+k_3/y^6)^{-2/3}dx^{\mu}dx_{\mu}+
(1+k_3/y^6)^{1/3}(dy^2+y^2d\Omega_7{}^2)
\ee
and the four-form field strength by
\be
\tilde K_7 \equiv *K_4=6k_3\epsilon_7
\ee
where the constant $k_3$ is given by
\be
k_3=\frac{2\kappa_{11}{}^2T_3}{\Omega_7}
\ee
Here $\epsilon_7$ is the volume form on $S^7$ and $\Omega_7$ is the volume. 
The mass 
per unit area of the membrane ${\cal M}_3$ is equal to its tension: 
\be
{\cal M}_3=T_3
\ee
This {\it elementary} solution is a singular solution of the supergravity 
equations coupled to a supermembrane source and carries a Noether
``electric'' charge  
\be
Q=\frac{1}{\sqrt{2}\kappa_{11}}\int_{S^7}(*K_4 + C_3 \wedge K_4)
=\sqrt{2}\kappa_{11}T_3 
\ee
Hence the solution saturates the Bogomol'nyi bound
$\sqrt{2}\kappa_{11}{\cal M}_3\geq Q$. This is a consequence of the
preservation of half the supersymmetries  which is also intimately linked
with the worldvolume kappa symmetry. The zero modes of this solution
belong to a $(d=3,n=8)$ supermultiplet consisting of eight scalars and
eight spinors $(\phi^I,\chi^I)$, with $I=1,...,8$, which correspond to the
eight Goldstone bosons and their superpartners associated with breaking of
the eight translations transverse to the membrane worldvolume.  

In \cite{Gueven}, the superfivebrane was discovered as a soliton solution
of $D=11$  supergravity also preserving half the spacetime supersymmetry
. Making the six/five split $X^M=(x^{\mu},y^m)$ where
$\mu=0,1,2,3,4,5$ and $m=6,...,10$, the metric is given by
\be
ds^2=(1+k_6/y^3)^{-1/3}dx^{\mu}dx_{\mu}+
(1+k_6/y^3)^{2/3}(dy^2+y^2d\Omega_4{}^2)
\ee
and the four-index field-strength by
\be
K_4=3k_6\epsilon_4
\ee
where the fivebrane tension ${\tilde T}_6$ is related to the constant $k_6$
by
\be
k_6=\frac{2\kappa_{11}{}^2{\tilde T}_6}{3\Omega_4}
\ee
Here $\epsilon_4$ is the volume form on $S^4$ and $\Omega_4$ is the volume.
The mass per unit $5$-volume of the fivebrane ${\cal M}_6$ is equal to its 
tension:
\be
{\cal M}_6={\tilde T}_6
\ee
This {\it solitonic} solution is a non-singular solution of the source-free
equations 
and carries a topological ``magnetic'' charge   
\be 
P=\frac{1}{\sqrt{2}\kappa_{11}}\int_{S^4}K_4=\sqrt{2}\kappa_{11}{\tilde
T}_6 
\ee
Hence the solution saturates the Bogomol'nyi
bound $\sqrt{2}\kappa_{11}{\cal M}_6\geq P$. Once again, this is a
consequence of the preservation of half the  supersymmetries. The
kappa covariant action (or even field equations) for this $D=11$
superfivebrane is still unknown (see \cite{Ortin,Howesezginp} for recent
progress) but consideration  of the soliton zero modes
\cite{Gibbonstownsend,Khuristring,Townsendeleven} means that the gauged fixed
action must be described by the same chiral antisymmetric tensor multiplet
$(B^-{}_{\mu\nu},\lambda^I,\phi^{[IJ]})$ as that of the Type $IIA$ fivebrane
\cite{Callan1,Callan2}. Note that in addition to the five scalars
corresponding to the five translational Goldstone bosons, there is also a
$2$-form $B^-{}_{\mu\nu}$ whose $3$-form field strength is anti-self dual and
which describes three degrees of freedom.    

The electric and magnetic charges obey a Dirac quantization rule
\cite{Nepomechie,Teitelboim} 
\be
QP=2\pi n \qquad n={\rm integer}
\ee
Or, in terms of the tensions \cite{Luelem,Lublack},
\begin{equation}
2\kappa_{11}{}^2 T_3 {\tilde T}_6 =2\pi  n
\label{Dirac11}
\end{equation}
This naturally suggests a $D=11$ membrane/fivebrane duality. Note that
this reduces the three dimensionful parameters $T_3$, ${\tilde T}_6$ and
$\kappa_{11}$ to two.   Moreover, it was recently shown
\cite{Duffliuminasian} that they are not independent.  To see this, we note
from (2.2) that $C_3$ has period $2\pi/T_3$ so that $K_4$ is quantized
according to   %
\begin{equation} \int K_4={2\pi n\over T_3}\,\,\,\,\,n=integer \label{kquant}
\end{equation}
Consistency of such $C_3$ periods with the spacetime action,
(\ref{supergravity11}), gives the relation
\begin{equation}
{(2\pi)^2\over\kappa_{11}{}^2T_3^3}\in 2Z
\label{eq:k11t3}
\end{equation}
From (\ref{Dirac11}), this may also be written as 
\begin{equation}
2\pi {\tilde T_6\over T_3{}^2}\in Z
\label{eq:newdirac}
\end{equation}
Thus the tension of the singly charged fivebrane is given by
\be
\tilde T_6=\frac{1}{2\pi}T_3{}^2
\la{tension}
\ee
in agreement \cite{Dealwis} with the $D$-brane derivation
\cite{Schwarzpower}. 

Having obtained these solitons in ten and eleven dimensions, a bewildering
array of other solitonic $p$-branes may be obtained by vertical and
diagonal dimensional reduction \cite{Khuristring,Stain,Brane}

\section{Duality}
\label{dual}

\subsection{String/fivebrane duality: good news and bad}
\label{fivebrane}

The $3$-form version of $D = 10, N =1$ supergravity admits the elementary
string as a solution \cite{Dabholkar}. Here we shall show that the
$7$-form version admits the elementary fivebrane as a solution
\cite{Luelem}. The elementary string can also be interpreted as a
magnetic-like solution of the $7$-form version, as can the elementary
fivebrane for the $3$-form version.  This suggests that the heterotic string
and the heterotic fivebrane may be dual to each other in the sense that they
are equivalent descriptions of the same underlying physical theory. In the
dual formulation, the metric $g_{MN}$ and the dilaton $\phi$ are the same,
whereas the $7$-form field strength $K$ of the fivebrane is dual to the
$3$-form field strength of the string $H$.  More precisely, 
\begin{equation}
{H = e^\phi {^\ast} K ,}
\end{equation} 
where $^\ast$ denotes the Hodge dual using the canonical metric, so that the
field equation of the $3$-form becomes the Bianchi identity of the $7$-form
and vice-versa. Hence
\begin{equation}
{I_{10}({\rm fivebrane}) = {1\over 2\kappa^2} \int d^{10} x
\sqrt{- g}
\big(R - {1\over 2 } (\partial \phi)^2 - {1\over 2\cdot 7!}
 e^\phi K^2 \big),}
\end{equation}
where $K$ is the curl of the 6-form $A$:
\begin{equation}
{K = d A.}
\end{equation}
String/fivebrane duality will tell us that
the fivebrane loop coupling constant ${\rm g}_6$ is given by
the inverse cube
root of the string loop coupling constant ${\rm g}_2$,
\begin{equation}{{\rm g}_6 = {\rm g}_2^{-1/3}=e^{-\phi_0/3}, }
\label{coupling}
\end{equation}
which implies that the strongly coupled heterotic string
corresponds to the
weakly coupled fivebrane, and vice versa. We will now derive
(\ref{coupling}) and the
following relations between the canonical gravitational metric
and the metrics
which appear naturally in
the string and fivebrane $\sigma$-models:
\begin{equation}{g_{MN} ({\rm canonical}) = e^{-\phi/2} g_{MN}({\rm
 string}) =
e^{\phi/6} g_{MN}({\rm fivebrane}).}
\label{metrics}
\end{equation}
In general, we have
\begin{equation}{g_{MN} ({\rm string}) = \Omega_s (\phi) g_{MN}
({\rm canonical}),}\end{equation}
and
\begin{equation}{g_{MN} ({\rm fivebrane} ) = \Omega_f (\phi) g_{MN}
({\rm canonical}). }\end{equation}
The corresponding lowest-order $\sigma$-model actions
written in terms of the canonical metric are given by
\begin{equation}{S_2 = - T_2 \int d^2 \xi ({1\over 2} \sqrt {-\gamma}
\gamma^{ij}
\partial_i X^M \partial_j X^N \Omega_s(\phi) g_{MN} + {1\over 2}
\varepsilon^{ij}\partial_i X^M \partial_j X^N B_{MN} ),}\end{equation}
for the string and
\[S_6 = - T_6 \int d^6 \xi \big({1\over 2}\sqrt {-\gamma}
\gamma^{ij}\partial_i X^M \partial_j X^N \Omega_f(\phi) g_{MN}
 - 2\sqrt {-\gamma}
\]
\begin{equation}
 +{1\over 6!} \varepsilon^{ijklmn} \partial_i X^M
\partial_j X^N
\partial_k X^O\partial_l X^P \partial_m X^Q
\partial_n X^R A_{MNOPQR} \big)
\end{equation}
for the fivebrane. Note that in this case
\begin{equation}{\gamma_{ij} = \partial_i X^M
\partial_j X^N \Omega_f (\phi) g_{MN}.}
\end{equation}
To fix $\Omega_s$ and $\Omega_f$, consider the following two parameter
constant rescalings
\begin{equation}
{g_{MN} \rightarrow {\lambda}^{1/2}{\sigma}^{3/2}\ g_{MN},}
\end{equation}
\begin{equation}
{e^\phi \rightarrow {\lambda}^{3}{\sigma}^{-3} e^\phi,}
\end{equation}
\begin{equation} 
{B_{MN} \rightarrow \lambda^2 B_{MN},}
\end{equation} 
\begin{equation}
{A_{MNOPQR} \rightarrow \sigma^6 A_{MNOPQR}.}
\end{equation}
 Provided we choose
\begin{equation}{\Omega_s (\phi) = e^{\phi/2}, \qquad \qquad
\Omega_f (\phi) =
e^{- \phi/6},}
\end{equation}
then the actions have the required homogeneous scaling
\begin{equation}
{ I_{10}({\rm string/fivebrane}) \rightarrow
{\lambda}^{2}{\sigma}^{-6}I_{10}({\rm string/fivebrane}),} 
\end{equation}
\begin{equation}
{S_2 \rightarrow \lambda^2 S_2 ,}
\end{equation}
\begin{equation}
{S_6 \rightarrow \sigma^6 S_6 ,}
\end{equation}
This result agrees with what one obtains in string theory by setting the
$\beta$-functions of $g_{MN}$, $B_{MN}$ and $\phi$ to zero, i.e. by using
string worldsheet conformal invariance.

We are now ready to derive (\ref{coupling}). This can be achieved
simply by writing
$I_{10}({\rm string})$ and $I_{10}({\rm fivebrane})$ in terms
of string and fivebrane metrics given in
(\ref{metrics}), respectively. We obtain for the string action
\begin{equation}
{I_{10}({\rm string}) = {1\over 2\kappa^2}
\int d^{10} x \sqrt {-g}
e^{-2\phi} \big(R + 4 (\partial\phi)^2 - {1\over 2\cdot 3!}
 H^2 \big),}
\end{equation}
where the common factor $e^{-2\phi}$ for each term implies
that the string loop counting parameter ${\rm g}_2$ is given
by
\begin{equation}
{{\rm g}_2 = e^{\phi_0}.}
\end{equation}
For the fivebrane action,
\begin{equation}{I_{10}({\rm fivebrane}) = {1\over 2 \kappa^2}
 \int d^{10} x
\sqrt {-g}
e^{2\phi/3}\big(R - {1\over 2\cdot 7!} K^2 \big),}
\end{equation}
and an analogous situation arises, namely a common factor
$e^{2\phi/3}$ in
front of each term. In analogy with the case of the string, the
common factor suggests that the fivebrane loop counting
parameter ${\rm g}_6$
is
given by
\begin{equation}{{\rm g}_6 = e^{-\phi_0/3}.}\end{equation}

String/fivebrane duality was now much more closely mimicking the
electric/magnetic duality of Montonen and Olive \cite{Montonen}.  However,
since most physicists were already sceptical of electric/magnetic duality
in four dimensions, they did not immediately embrace string/fivebrane
duality in ten dimensions! 

Furthermore, there was one major problem with treating the fivebrane as
a fundamental object in its own right; a problem that has bedevilled
supermembrane theory right from the beginning: no-one knows how to
quantize fundamental $p$-branes with $p>2$.  All the techniques that
worked so well for fundamental strings and which allow us, for example, to
calculate how one string scatters off another, simply do not go through.
Problems arise both at the level of the worldvolume equations where the 
{\it bete noir} of non-renormalizability comes to haunt us and also at the
level of the spacetime equations. Each term in string perturbation theory
corresponds to a two-dimensional worldsheet with more and more holes: we
must sum over all topologies of the worldsheet. But for surfaces with more
than two dimensions we do not know how to do this.  Indeed, there are
powerful theorems in pure mathematics which tell you that it is not merely
hard but impossible. In any case, the quantities that are light in the
limit of weak coupling are particles and strings but not branes with $p
\geq 2$ \cite{Hull}. So there were two major impediments to
string/fivebrane duality in $10$ dimensions. First, the electric/magnetic
duality analogy was ineffective so long as most physicists were sceptical
of this duality. Secondly, treating fivebranes as fundamental raised all
the unresolved issues of quantization.  

\subsection{String/string duality in $D=6$}
\label{stringduality}

The first of these impediments was removed, however, when Sen \cite{SenS}
revitalized the Olive-Montonen conjecture by establishing that certain
dyonic states, which their conjecture demanded, were indeed present in the
theory. Many duality sceptics were thus converted.  Indeed this inspired
Seiberg and Witten \cite{Seiberg} to look for duality in more
realistic (though still supersymmetric) approximations to the standard
model. The subsequent industry, known as Seiberg-Witten theory, provided a
wealth of new information on non-perturbative effects in four-dimensional
quantum field theories, such as quark-confinement and symmetry-breaking,
which would have been unthinkable just a few years ago.
  
As for the second problem of fundamental fivebranes, recall that when
wrapped around a circle, an $11$-dimensional membrane behaves as if it
were a $10$-dimensional string. In a series of papers
\cite{Luloop,Khurifour,Lublack,Minasian1,Duffrahmfeld,Khuristring,%
Duffstrong,Duffliuminasian}
between 1991 and 1995, Duff, Khuri, Liu, Lu, Minasian and Rahmfeld argued
that this may also be the way out of the problems of $10$-dimensional
string/fivebrane duality. If we allow four of the ten dimensions to be
curled up and allow the solitonic fivebrane to wrap around them, it will
behave as if it were a $6$-dimensional  solitonic string! The fundamental
string will remain a fundamental string but now also in $6$-dimensions. So
the $10$-dimensional string/fivebrane duality conjecture gets replaced by
a $6$-dimensional string/string duality conjecture. The obvious advantage
is that, in contrast to the fivebrane, we do know how to quantize the
string and hence we can put the predictions of string/string duality to
the test. For example, one can show that the coupling constant of the
solitonic string is indeed given by the inverse of the fundamental
string's coupling constant, in complete agreement with the conjecture. 

When we spoke of string/string duality, we originally had in mind a
duality between one heterotic string and another, but the next major
development in the subject came in 1994 when Hull and Townsend
\cite{Hulltownsend} suggested that, if the four-dimensional compact space
is chosen suitably, a six-dimensional heterotic string can be dual to a
six-dimensional Type $IIA$ string!    

Evidence in favor of the idea that the physics of the
fundamental string in six spacetime dimensions may equally well be
described by a dual string and that one emerges as a soliton solution of
the other
\cite{Luloop,Lublack,Minasian1,Duffstrong,Sensol,Harveystrominger} goes as
follows. The string equations admits the singular {\it elementary} string
solution \cite{Dabholkar}    
\[
ds^2= (1-k^2/r^2)[-d\tau^2+d\sigma^2 + (1-k^2/r^2)^{-2}dr^2
+r^2d\Omega_{3}{}^2]
\]
\[
e^{\Phi}=1-k^2/r^2
\]
\be
e^{-\Phi}*H_3=2k^2\epsilon_3
\la{fundstring}
\ee
where
\be
k^2=\kappa^2 T/\Omega_3
\ee
$T=1/2\pi \alpha'$ is the string tension, $\Omega_3$ is the volume of $S^3$ and
$\epsilon_3$ is the volume form. It describes an
infinitely long string whose worldsheet lies in the plane $X^0=\tau,X^1
=\sigma$.  Its mass per unit length is given by 
\be
 M= T<e^{\Phi/2}>
\ee
and is thus heavier for stronger string coupling, as one would expect for
a fundamental string.  The string equations also admit the non-singular
{\it solitonic} string solution
\cite{Lublack,Minasian1}
\[
ds^2= -d\tau^2+d\sigma^2 + (1-\tilde k^2/r^2)^{-2}dr^2 + r^2d\Omega_{3}{}^2
\]
\[
e^{-\Phi}=1-\tilde k^2/r^2
\]
\be
H_3=2\tilde k^2\epsilon_3
\la{sol}
\ee
whose tension $\tilde T=1/2\pi {\tilde \alpha}'$ is given by
\be
\tilde k^2=\kappa^2 \tilde T/\Omega_3
\ee
Its mass per unit length is given by
\be
\tilde {M}= \tilde T <e^{-\Phi/2}>
\ee
and is thus heavier for weaker string coupling, as one would expect
for a solitonic string. Thus we see that the solitonic string differs from the
fundamental string by the replacements 
\[
\Phi \rightarrow \tilde \Phi=-\Phi
\]
\[
G_{MN} \rightarrow \tilde G_{MN}=e^{-\Phi}G_{MN}
\] 
\[
H \rightarrow \tilde H=e^{-\Phi}*H
\]
\be
\alpha' \rightarrow \tilde \alpha'
\la{dualstring}
\ee
The Dirac quantization rule $eg=2\pi n$ ($n$=integer) relating the Noether
``electric'' charge
\be
e=\frac{1}{\sqrt{2}\kappa}\int_{S^3}e^{-\Phi}*H_3
\ee
to the topological ``magnetic'' charge
\be
g=\frac{1}{\sqrt{2}\kappa}\int_{S^3}H_3
\ee
translates into a quantization condition on the two tensions:
\be
2\kappa{}^2=n(2\pi)^3\alpha'\tilde \alpha'\,\,\,\,\,\,n=integer
\la{Dirac1}
\ee
where $\kappa$ is the six-dimensional gravitational constant. (However,
one may, by siutable choice of units, set the two tensions equal
\cite{DMW}.) Both the string and dual string soliton solutions break half
the supersymmetries, both saturate a Bogomol'nyi bound between the mass and
the charge. These solutions are the extreme mass equals
charge limit of more general two-parameter black string solutions
\cite{Horowitz1,Lublack}.

We now make the major assumption of string/string duality: the dual
string may be regarded as a fundamental string in its own right with a 
worldsheet action that couples to the dual variables and has the dual
tension given in (\ref{dualstring}). It follows that the dual string
equations admit the dual string (\ref{sol}) as the fundamental solution
and the fundamental string (\ref{fundstring}) as the dual solution.	When
expressed  in terms of the dual metric, however, the former is singular
and the latter non-singular. It also follows from (\ref{Dirac1}) that in
going from the fundamental string to the dual string and interchanging 
$\alpha'$ with  ${\tilde\alpha}'=2\kappa^2/(2\pi)^3\alpha'$, one also
interchanges the  roles of worldsheet and spacetime loop expansions.
Moreover, since the dilaton enters the dual string equations with the
opposite sign to the fundamental string, it was argued in
\cite{Luloop,Lublack,Minasian1} that in $D=6$ the strong coupling regime
of the string should correspond to the weak coupling regime of the dual
string:  %
\be
{\rm g}_6{}^2/(2\pi)^3 = <e^{\Phi}>=(2\pi)^3/{\tilde{\rm g}_6}^2
\la{stringcoupling}
\ee
where ${\rm g_6}$ and $\tilde{\rm g}_6$ are the six-dimensional string
and dual string loop expansion parameters.

\section{Electric/Magnetic Duality from String/String Duality}
\label{strong}

\subsection{ $S$ duality and $T$ duality}
\label{SandT}

In 1977 Montonen and Olive \cite{Montonen}, made a bold
conjecture. Might there exist a {\it dual} formulation of fundamental
physics in which the roles of Noether charges and topological charges are
reversed? In such a dual picture, the magnetic monopoles would be the
fundamental objects and the quarks, W-bosons and Higgs particles would be
the solitons! They were inspired by the observation that in certain {\it
supersymmetric} gauge theories, the masses $M$ of all the particles
whether elementary (carrying purely electric charge $Q$), solitonic
(carrying purely magnetic charge $P$) or dyonic (carrying both) are
described by a universal formula 
\be
M^2=v^2(Q^2+P^2)
\ee
where $v$ is a constant. Note that the mass formula remains unchanged if
we exchange the roles of $P$ and $Q$! The Montonen-Olive conjecture was
that this electric/magnetic symmetry is a symmetry not merely of the mass
formula but is an exact symmetry of the entire quantum theory! The
reason why this idea remained merely a conjecture rather than a proof
has to do with the whole question of perturbative versus non-perturbative
effects. According to Dirac, the electric charge $Q$ is quantized in units
of $e$, the charge on the electron, whereas the magnetic charge
is quantized in units of $1/e$. In other words, $Q=me$ and $P=n/e$, where
$m$ and $n$ are integers. The symmetry suggested by Olive and Montonen
thus demanded that in the dual world, we not only exchange the integers
$m$ and $n$ but we also replace $e$ by $1/e$ and go from a regime of weak
coupling to a regime of strong coupling! 

The Olive-Montonen conjecture was originally intended to apply to
four-dimensional grand unified field theories. In 1990, however, 
Font, Ibanez, Lust and Quevedo \cite{Font} and,
independently, Rey \cite{Rey} generalized the idea to four-dimensional
superstrings, where in fact the idea becomes even more natural and goes by the
name of $S$-duality.  In fact, superstring theorists had already become used 
to a totally different kind of duality called $T$-duality \cite{Giveon}.
Unlike, $S$-duality which was a non-perturbative symmetry and hence still
speculative, $T$-duality was a perturbative symmetry and rigorously
established. If we compactify a string theory on a circle then,
in addition to the Kaluza-Klein particles we would expect in an ordinary
field theory, there are also extra {\it winding}
particles that arise because a string can wind around the circle.
$T$-duality states that nothing changes if we exchange the roles of the
Kaluza-Klein and winding particles provided we also exchange the radius
of the circle $R$ by its inverse $1/R$. In short, a string cannnot tell
the difference between a big circle and a small one!

\subsection{Duality of dualities}
\label{dualdual}

String/string duality now  has another unexpected pay-off
\cite{Duffstrong}. If we compactify the six-dimensional spacetime on two
circles down to four dimensions, the fundamental string and the solitonic
string will each acquire a $T$-duality. But here is the miracle: the
$T$-duality of the solitonic string is just the $S$-duality of the
fundamental string, and vice-versa! This phenomenon, in which the
non-perturbative replacement of $e$ by $1/e$ in one picture is just the
perturbative replacement of $R$ by $1/R$ in the dual picture, goes by the
name of {\it Duality of Dualities}.  Thus four-dimensional
electric/magnetic duality, which was previously only a conjecture, now
emerges automatically if we make the more primitive conjecture of
six-dimensional string/string duality.

To see this, we begin by recalling that when we include a
non-vanishing theta angle, the $S$-duality is desribed by $SL(2,Z)$. The
conjectured $SL(2,Z)$ symmetry  acts on the gauge coupling constant
$e$ and theta angle $\theta$ via: 
\be
S \rightarrow \frac{aS+b}{cS+d}
\la{sl2zs}
\ee
where $a,b,c,d$ are integers satisfying $ad-bc=1$ and where
\be
S=S_1+iS_2=\frac{\theta}{2\pi} + i\frac{4\pi}{e^2}
\la{S}
\ee
This is also called electric/magnetic duality because the integers $m$ and
$n$ which characterize the magnetic charges $Q_m=n/e$ and electric charges
$Q_e=e(m+n\theta/2\pi)$ of the particle spectrum transform as
\be
\left( \begin{array}{c}
m\\
n
\end{array}
\right)
\rightarrow
\left( \begin{array}{cc}
a&b\\
c&d
\end{array}
\right)
\left( \begin{array}{c}
m\\
n
\end{array}
\right)
\la{charges}
\ee
Such a symmetry would be inherently non-perturbative since, for $\theta=0$
and with $a=d=0$ and $b=-c=-1$, it reduces to the strong/weak coupling duality  
\[
{e}^2/4\pi \rightarrow4\pi/{e}^2
\]
\be
n\rightarrow m, m\rightarrow -n
\label{simple}
\ee
This in turn means that the coupling constant cannot get renormalized in
perturbation theory and hence that the renormalization group
$\beta$-function vanishes
\be
\beta(e)=0
\ee
This is guaranteed in $N=4$ supersymmetric Yang-Mills and also happens in
certain $N=2$ theories.  Thus the duality idea is that the theory
may equivalently be described in one of two ways. In the conventional way, the
W-bosons, Higgs bosons and their fermionic partners are the
electrically charged elementary particles and the magnetic monopoles emerge
as soliton solutions of the field equations.  In the dual description, however,
it is the monopoles which are elementary and the electrically charged particles
which emerge as the solitons.

In string theory the roles of the theta angle $\theta$ and coupling
constant $e$ are played by the VEVs of the the four-dimensional axion
field $a$ and dilaton field $\eta$:  
\be
<a>=\frac{\theta}{2\pi}
\ee
\be
{e}^2/4\pi=<e^{\eta}>=8G/\alpha'
\ee
Here $G$ is Newton's constant and $2\pi\alpha'$ is the inverse string tension.
Hence $S$-duality (\ref{sl2zs}) now becomes a transformation law for the
axion/dilaton field $S$: 
\be
S=S_1+iS_2={\rm a}+ie^{-\eta}
\la{Sfield}
\ee

Let us first consider $T$-duality and focus just on the moduli fields that arise
in  compactification on a $2$-torus of a $D=6$ string with dilaton $\Phi$, 
metric $G_{MN}$ and $2$-form potential $B_{MN}$ with $3$-form field
strength $H_{MNP}$. Here the $T$-duality is just $O(2,2;Z)$.  Let us
parametrize the compactified ($m,n=4,5$) components of string metric and
2-form as    
\be
G_{mn}=e^{\rho-\sigma}\left( \begin{array}{cc}
e^{-2\rho}+c^2&-c\\
-c&1
\end{array}\right)
\ee
and
\be
B_{mn}=b\epsilon_{mn}
\ee
The four-dimensional shifted dilaton $\eta$ is given by   
\be
e^{-\eta}=e^{-\Phi}\sqrt{det G_{mn}}=e^{-\Phi -\sigma}
\ee
and the axion field ${\rm a}$ is defined by 
\be
\epsilon^{\mu\nu\rho\sigma}\partial_{\sigma}{\rm a}=
\sqrt{-g}e^{-\eta}g^{\mu\sigma}g^{\nu\lambda}g^{\rho\tau}H_{\sigma\lambda\tau}
\ee
where $g_{\mu\nu}=G_{\mu\nu}$ and $\mu,\nu=0,1,2,3$. We further define the 
complex
Kahler form field $T$ and the complex structure field $U$ by  
\[
T=T_1+iT_2=b+ie^{-\sigma}
\]
\be
U=U_1+iU_2=c+ie^{-\rho}
\ee
Thus this $T$-duality may be written as
\be
O(2,2;Z)_{TU} \sim SL(2,Z)_T \times SL(2,Z)_U
\la{Tsplit}
\ee
where $SL(2,Z)_T$ acts on the $T$-field and $SL(2,Z)_U$ acts on the $U$-field
in the same way that $SL(2,Z)_S$ acts on the $S$-field in (\ref{sl2zs}).  In
contrast to $SL(2,Z)_S$, $SL(2,Z)_T \times SL(2,Z)_U$ is known to be
an exact string symmetry order by order in string perturbation theory.
$SL(2,Z)_T$ does, however, contain a minimum/maximum length duality
mathematically similar to (\ref{simple})   
\be
R \rightarrow \alpha'/R
\la{R}
\ee
where $R$ is the compactification scale given by
\be
\alpha'/R^2=<e^{\sigma}>.
\ee
Let us now investigate how both $N=4$ and $N=2$ exact electric/magnetic
duality follows from string theory.  There is a formal
similarity between $S$-duality and $T$-duality.  It was argued in
\cite{Duffstrong} that these mathematical similarities between $SL(2,Z)_S$
and $SL(2,Z)_T$ are not coincidental. On compactification to four
spacetime dimensions, the two theories appear very similar, each acquiring
an $O(2,2;Z)$ target space duality.	One's first guess might be to assume
that the strongly coupled four-dimensional fundamental
string corresponds to the weakly coupled dual string, but in fact
something more subtle and interesting happens: the roles of the $S$ and $T$
fields are interchanged \cite{Khurifour} so that the strong/weak coupling
$SL(2,Z)_S$ of the fundamental string emerges as a subgroup of the target
space duality of the dual string:
\be
O(2,2;Z)_{SU} \sim SL(2,Z)_S \times SL(2,Z)_U
\la{Ssplit}
\ee
This {\it duality of dualities} is summarized in Table \ref{table}.
\begin{table}
\caption{Duality of dualities}
\label{table}
$
\begin{array}{lll}
&Fundamental \, string&Dual \, string\\
&&\\
T-duality&O(2,2;Z)_{TU} &O(2,2;Z)_{SU}\\
&\sim SL(2,Z)_T \times	SL(2,Z)_U&\sim SL(2,Z)_S\times	SL(2,Z)_U\\
Moduli&T={\rm b}+ie^{-\sigma}&S={\rm a}+ie^{-\eta}\\
&{\rm b}=B_{45}&{\rm a}=\tilde B_{45}\\
&e^{-\sigma}=\sqrt{detG_{mn}}&e^{-\eta}=\sqrt{det \tilde{G}_{mn}}\\
Worldsheet \, coupling&<e^{\sigma}>=\alpha'/R^2&<e^{\eta}>={\rm g}^2/2\pi\\
Large/small \, radius &R\rightarrow \alpha'/R&{\rm g}^2/2\pi\rightarrow
2\pi/{\rm g}^2\\
S-duality&SL(2,Z)_S&SL(2,Z)_T\\
Axion/dilaton&S={\rm a}+ie^{-\eta}&T={\rm b}+ie^{-\sigma}\\
&d{\rm a}=e^{-\eta}*H&d{\rm b}=e^{-\sigma}\tilde{*} \tilde{H}\\
&e^{-\eta}=e^{-\Phi}\sqrt{detG_{mn}}&e^{-\sigma}=e^{\Phi}\sqrt{det
\tilde{G}_{mn}}\\
Spacetime \, coupling&<e^{\eta}>={\rm g}^2/2\pi&<e^{\sigma}>=\alpha'/R^2\\
Strong/weak \, coupling&{\rm g}^2/2\pi\rightarrow 2\pi/{\rm g}^2&R
\rightarrow \alpha'/R
\end{array}
$
\end{table}
As a consistency check, we note that since $(2\pi R)^2/2\kappa^2=1/16\pi G$
the Dirac
quantization rule (\ref{Dirac1}) becomes (choosing $n$=1)
\be
8GR^2=\alpha'\tilde \alpha'
\la{Dirac2}
\ee
Invariance of this rule now requires that a strong-weak coupling
transformation on the fundamental
string ($8G/\alpha'\rightarrow \alpha'/8G$) must be accompanied by a
minimum/maximum length
transformation of the dual string ($\tilde \alpha'/R^2 \rightarrow
R^2/\tilde \alpha'$), and vice
versa.

\section{String/String Duality from M Theory}
\label{duality}

All this previous work on $T$-duality, $S$-duality, and string/string
duality was suddenly pulled together by Witten \cite{Wittenvarious} under the
umbrella of eleven-dimensions.  One of the biggest problems  with $D=10$
string theory is that there are {\it five} consistent string
theories:  Type $I$ $SO(32)$, heterotic  $SO(32)$, heterotic $E_8 \times
E_8$,  Type $IIA$ and Type $IIB$. As a candidate for a unique {\it theory
of everything}, this is clearly an embarrassment of riches.  Witten put
forward a convincing case that this distinction is just an artifact of
perturbation theory and that non-perturbatively these five theories are,
in fact, just different corners of a deeper theory. Moreover, this deeper
theory, subsequently dubbed {\it $M$-theory}, has $D=11$ supergravity as
its low energy limit! Thus the five string theories and $D=11$
supergravity represent six different special points\footnote{Some authors
take the phrase {\it $M$-theory} to refer merely to this sixth corner of
the moduli space. With this definition, of course, $M$-theory is no more
fundamental than the other five corners. For us, {\it $M$-theory} means
the whole kit and caboodle.} in the moduli space of $M$-theory.  The small
parameters of perturbative string theory are provided by $<e^{\Phi}>$,
where $\Phi$ is the dilaton field, and $<e^{{\sigma}_i}>$ where
${\sigma}_i$ are the moduli fields which arise after compactification.
What makes $M$-theory at once intriguing and yet difficult to analyse is
that in $D=11$ there is neither dilaton nor moduli and hence the theory is
intrinsically non-perturbative. Consequently, the ultimate meaning of
$M$-theory is still unclear, and Witten has suggested that in the
meantime, $M$ should stand for ``Magic'', ``Mystery'' or ``Membrane'',
according to taste. Curiously enough, however, Witten still played down
the importance of supermembranes. But it was only a matter of time before
he too succumbed to the conclusion that we weren't doing just string
theory any more! In the coming months, literally hundreds of papers
appeared in the internet confirming that, whatever $M$-theory may be, it
certainly involves supermembranes in an important way. For example, we
shall now see how the $6$-dimensional string/string duality of section
(\ref{stringduality}) (and hence the $4$-dimensional electric/magnetic
duality of section (\ref{dualdual})) follows from $11$-dimensional
membrane/fivebrane duality \cite{Duffliurahmfeld,DMW}. The fundamental
string is obtained by wrapping the membrane around a one-dimensional space
and then compactifying on a four-dimensional space; whereas the solitonic
string is obtained by wrapping the fivebrane around the four-dimensional
space and then compactifying on the one-dimensional space. Nor did it take
long before the more realistic kinds of electric/magnetic duality
envisioned by Seiberg and Witten \cite{Seiberg} were also given an
explanation in terms of string/string duality and hence $M$-theory.

Even the chiral $E_8 \times E_8$ string, which according to Witten's
earlier theorem could never come from eleven-dimensions, was given an
eleven-dimensional explanation by Horava and Witten \cite{Horava}. 
The no-go theorem is evaded by compactifying not on a circle (which has no
ends), but on a line-segment (which has two ends).  Witten went on to argue
that if the size of this one-dimensional space is large compared to the
six-dimensional Calabi-Yau manifold, then our world is approximately
five-dimensional. This may have important consequences for confronting
$M$-theory with experiment. For example, it is known that the strengths of
the four forces change with energy. In supersymmetric extensions of the
standard model, one finds that the fine structure constants
$\alpha_3,\alpha_2,\alpha_1$ associated with the $SU(3) \times SU(2)
\times U(1)$  all meet at about $10^{16}$ GeV, entirely consistent with
the idea of grand unification.  The strength of the dimensionless number
$\alpha_G=GE^2$, where $G$ is Newton's contant and $E$ is the energy, also
almost meets the other three, but not quite. This near miss has been a
source of great interest, but also frustration.  However, in a universe of
the kind envisioned by Witten \cite{Wittencalabi}, spacetime is
approximately a narrow five dimensional layer bounded by four-dimensional
walls. The particles of the standard model live on the walls but gravity
lives in the five-dimensional bulk. As a result, it is possible to  choose
the size of this fifth dimension so that all four forces meet at this
common scale.  Note that this is much less than the Planck scale of
$10^{19}$ GeV, so gravitational effects may be much closer in energy than
we previously thought; a result that would have all kinds of cosmological
consequences.

Thus this eleven-dimensional framework now provides the starting point for
understanding a wealth of new non-perturbative phenomena, including
string/string duality, Seiberg-Witten theory, quark confinement,
particle physics phenomenology and cosmology.

\subsection{Membrane/fivebrane duality in $D=11$}
\label{Eleven}

Let us consider $M$-theory, with its fundamental membrane and solitonic 
fivebrane, on $R^6 \times M_1 \times {\tilde M_4}$ where $M_1$ is a
one-dimensional compact space of radius $R$ and ${\tilde M_4}$ is a
four-dimensional compact space of volume $V$. We may obtain a fundamental
string on $R^6$ by wrapping the membrane around $M_1$ and reducing on
${\tilde M_4}$. Let us denote fundamental string sigma-model metrics in
$D=10$ and $D=6$ by $G_{10}$ and $G_{6}$. Then from the corresponding
Einstein Lagrangians   
\be
\sqrt{-G_{11}}R_{11}=
R^{-3}\sqrt{-G_{10}}R_{10}=\frac{V}{R}\sqrt{-G_{6}}R_{6}
\ee
we may read off the strength of the string couplings in $D=10$ \cite{DMW}
\be
{\lambda_{10}}^2=R^3
\ee
and $D=6$
\be
{\lambda_{6}}^2=\frac{R}{V}
\ee
Similarly we may obtain a solitonic string on $R^6$ by wrapping the 
fivebrane around ${\tilde M_4}$ and reducing on $M_1$. Let us denote
the solitonic string sigma-model metrics in $D=7$ and $D=6$ by ${\tilde
G}_{7}$ and ${\tilde G}_{6}$. Then from the corresponding Einstein
Lagrangians   \be 
\sqrt{-G_{11}}R_{11}=
V^{-3/2}\sqrt{-{\tilde G}_{7}}{\tilde R}_{7}=
\frac{R}{V}\sqrt{-{\tilde G}_{6}}{\tilde R}_{6} 
\ee 
we may read off the strength of the string couplings in $D=7$ \cite{DMW} 
\be
{{\tilde \lambda}_{7}}^2=V^{3/2}
\ee
and $D=6$
\be
{{\tilde \lambda}_{6}}^2=\frac{V}{R}
\ee
Thus we see that the fundamental and solitonic strings are related by a 
strong/weak coupling:
\be
{{\tilde \lambda}_{6}}^2=1/{\lambda_{6}}^2
\ee

\begin{table}
\caption{String/string dualities}
\label{scenario}
$
\begin{array}{ccccc}
~~~~~~~~~~~~~~{\bf (N_+,N_-)}&{\bf M_1}&{\bf {\tilde
M}_4}&{\bf fundamental~string}&{\bf dual~string}\\ ~~~~~~~~~~~~~~&~&~&\\
~~~~~~~~~~~~~~(1,0)&S^1/Z_2&K3&heterotic&heterotic\\
~~~~~~~~~~~~~~(1,1)&S^1&K3&Type~IIA&heterotic\\
~~~~~~~~~~~~~~(1,1)&S^1/Z_2&T^4&heterotic&Type~IIA\\
~~~~~~~~~~~~~~(2,2)&S^1&T^4&Type~IIA&Type~IIA
\end{array}
$
\end{table}

We shall be interested in $M_1=S^1$ in which case the fundamental string
will be Type $IIA$ or $M_1=S^1/Z_2$ in which  case the fundamental string
will be heterotic $E_8 \times E_8$. Similarly, we will be interested in
${\tilde M}_4=T^4$ in which case the solitonic string will be Type $IIA$
or ${\tilde M}_4=K3$ in which case the solitonic string will be heterotic.
Thus there are four possible scenarios which are summarized in Table
\ref{scenario}. $(N_+,N_-)$ denotes the $D=6$ spacetime supersymmetries.
In each case, the fundamental string will be weakly coupled as we shrink
the size of the wrapping space $M_1$ and the dual string will be weakly
coupled as we shrink the size of the wrapping space ${\tilde M}_4$. 

In fact, there is in general a topological
obstruction to wrapping the fivebrane around  ${\tilde M}_4$ provided by
(\ref{kquant}) because the fivebrane cannot wrap around a $4$-manifold that
has $n\neq 0$\footnote{Actually,
as recently shown in \cite{Wittenflux}, the object which must have integral
periods is not $T_3K_4/2\pi$ but rather $T_3K_4/2\pi - p_1/4$ where $p_1$ is the
first Pontryagin class.  This will not affect our conclusions, however.}.  This
is because the anti-self-dual $3$-form field strength $T$ on the worldvolume of
the fivebrane obeys \cite{TownsendM,Wittenfive}   \be dT=K_4 \ee
and the existence of a solution for $T$ therefore requires that $K_4$ must
be cohomologically trivial. For $M$-theory on $R^6 \times S^1/Z_2 \times
T^4$ this is no problem. For $M$ theory on $R^6 \times S^1/Z_2 \times K3$,
with instanton number $k$ in one $E_8$ and $24-k$ in the other, however,
the flux of $K_4$ over $K3$ is 
\cite{DMW} 
\be 
n=12-k
\ee
Consequently, the $M$-theoretic explanation of heterotic/heterotic duality
requires $E_8 \times E_8$ with the symmetric embedding $k=12$. This
has some far-reaching implications. For example, the duality exchanges
gauge fields that can be seen in perturbation theory with gauge fields of a
non-perturbative origin \cite{DMW}.

The dilaton $\tilde \Phi$, the string
$\sigma$-model metric $\tilde G_{MN}$ and $3$-form field strength $\tilde
H$ of the dual string are related to those of the fundamental string,  $\Phi$,
$G_{MN}$ and $H$ by the replacements \cite{Lublack,Minasian1}      
\[
\Phi \rightarrow \tilde \Phi=-\Phi
\]
\[
G_{MN} \rightarrow \tilde G_{MN}=e^{-\Phi}G_{MN}
\] 
\be
H \rightarrow \tilde H=e^{-\Phi}*H
\la{discrete}
\ee
In the case of heterotic/Type $IIA$ duality and Type $IIA$/heterotic
duality, this operation takes us from one string to the other, but
in the case of heterotic/heterotic duality and Type $IIA$/Type $IIA$
duality this operation is a discrete symmetry of the theory. This 
Type $IIA$/Type $IIA$ duality is discussed in \cite{Senvafa}
and we recognize this symmetry as subgroup of the
$SO(5,5;Z)$ $U$-duality \cite{Luduality,Hulltownsend,Dijkgraaf} of the
$D=6$ Type $IIA$ string. 

Vacua with $(N_+,N_-)=(1,0)$ in $D=6$ have been the subject
of much interest lately. In addition to DMW vacua \cite{DMW}
discussed above, obtained from $M$-theory on $S^1/Z_2 \times K3$, there
are also the GP vacua \cite{Pradisi,Bianchi,Gimon} obtained from the $SO(32)$
theory on $K3$ and the MV vacua \cite{Morrison1,Morrison2} obtained from {\it
$F$-theory} \cite {Vafa} on Calabi-Yau.  Indeed, all three categories are
related by duality
\cite{Alda1,Morrison1,Gross,Wittenphase,Ferrara2,Berkooz,Morrison2,Alda2}.
In particular, the DMW heterotic strong/weak coupling duality gets mapped
to a $T$-duality of the Type $I$ version of the $SO(32)$ theory, and the
non-perturbative gauge symmetries of the DMW model arise from small
$Spin(32)/Z_2$ instantons in the heterotic version of the $SO(32)$ theory
\cite{Berkooz}.  Because heterotic/heterotic duality interchanges
worldsheet and spacetime loop expansions -- or because it acts by duality
on $H$ -- the duality exchanges the tree level Chern-Simons contributions
to the Bianchi identity \[ dH=\alpha'(2\pi)^2X_4  \] 
\be
X_4=\frac{1}{4(2\pi)^2}[\tr R^2-\Sigma_\alpha v_\alpha \tr F_\alpha{}^2] 
\la{Bianchi} 
\ee
with the  one-loop Green-Schwarz corrections to the field equations
\[
d\tilde H=\alpha'(2\pi)^2\tilde X_4
\]
\be
\tilde X_4=\frac{1}{4(2\pi)^2}[\tr R^2-\Sigma_\alpha {\tilde v}_\alpha 
\tr F_\alpha{}^2]
\la{field}
\ee
Here $F_\alpha$ is the field strength of the $\alpha^{th}$ component
of the gauge group, $\tr$ denotes the trace in the fundamental
representation, and $v_\alpha,\tilde v_\alpha$ are constants. (Without
loss of generality, we may choose the string and dual string tensions to
be equal \cite{DMW}.) In fact, the Green-Schwarz anomaly cancellation
mechanism in six dimensions requires that the anomaly eight-form $I_8$
factorize as a product of four-forms,   \be 
I_8=X_4\tilde X_4,
\la{eightform} 
\ee
and a six-dimensional string-string duality with the general features
summarized above would exchange the two factors \cite{Minasian1}. 
Moreover, supersymmetry relates the coefficients $v_\alpha,\tilde
v_\alpha$ to the gauge field kinetic energy.  In the Einstein  metric
$G^c{}_{MN}=e^{-\Phi/2}G{}_{MN}$, the exact dilaton dependence of the
kinetic energy of the gauge field $F_\alpha{}_{MN}$, is \cite{Sagnotti} 
\be 
L_{gauge}=-\frac{(2\pi)^3}{8\alpha'}\sqrt{G^c}\Sigma_\alpha \left( v_\alpha
e^{-\Phi/2}  +\tilde v_\alpha e^{\Phi/2} \right)\tr
F_\alpha{}_{MN}F_\alpha{}^{MN}. 
\la{happiness}
\ee
So whenever one of the $\tilde v_\alpha$ is negative, there is a value of
the dilaton for which the coupling constant of the corresponding gauge
group diverges. This is believed to signal a phase transition associated
with the appearance of tensionless strings
\cite{Ganor,Seibergwitten,Dufflupope}.  This does not happen for the
symmetric embedding discussed above since the perturbative gauge fields
have $v_\alpha>0$ and $\tilde v_\alpha=0$ and the non-perturbative gauge
fields have $v_\alpha=0$ and $\tilde v_\alpha>0$. Another kind of
heterotic/heterotic duality may arise, however, in vacua where one may
Higgs away that subset of gauge fields with negative $\tilde v_\alpha$,
and be left with gauge fields with $v_\alpha=\tilde v_\alpha>0$.  This
happens for the non-symmetric embedding $k=14$ and the appearance
of non-perturbative gauge fields is not required 
\cite{Alda1,Alda2,Johnson,Morrison1,Morrison2}.
Despite appearances, it known from $F$-theory that the $k=12$
and $k=14$ models are actually equivalent \cite{Morrison1,Morrison2}.

Vacua with $(N_+,N_-)=(2,0)$ arising from Type $IIB$ on $K3$ also have an
$M$-theoretic description, in terms of compactification on $T^5/Z_2$
\cite{Dasgupta,Wittenfive}.

\subsection{Four dimensions}
\label{four}

It is interesting to consider further toroidal compactification to four
dimensions, replacing $R^6$ by $R^4\times T^2$.  Starting with
a $K3$ vacuum in which the $E_8\times E_8$ gauge symmetry is completely
Higgsed, the toroidal compactification to four dimensions gives an
$N=2$ theory with the usual three vector multiplets $S$, $T$ and $U$ related
to the four-dimensional heterotic string coupling constant and
the area and shape of the $T^2$.  When reduced to four dimensions,
the six-dimensional string-string duality (\ref{discrete}) becomes
\cite{Duffstrong} an operation that exchanges $S$ and $T$, so in the case
of heterotic/heterotic duality we have a discrete $S-T$ interchange
symmetry.  This self-duality of heterotic string vacua does not rule out
the the possibility that in $D=4$ they are also dual to  Type $IIA$
strings compactified on  Calabi-Yau manifolds. In fact, as discussed in 
\cite{Kachru}, when the gauge group is completely Higgsed, obvious
candidates are provided by Calabi-Yau manifolds with hodge numbers
$h_{11}=3$ and $h_{21}=243$, since these have the same massless field
content. Moreover, these manifolds do indeed exhibit the $S-T$ interchange
symmetry \cite{Klemm,Duffelectric,Cardoso}. Since the heterotic string on
$T^2\times K3$ also has $R$ to $1/R$ symmetries that exchange $T$ and $U$,
one might expect a complete $S-T-U$ triality  symmetry, as discussed in
\cite{Duffliurahmfeld}. In all known models, however, the $T-U$
interchange symmetry is spoiled by non-perturbative effects
\cite{Louis,Cardoso1}.

An interesting aspect of the Calabi-Yau manifolds $X$ appearing in
the duality between heterotic strings on $K3 \times T^2$ and Type $IIA$
strings on $X$, is that they can always be written in the form of a $K3$
fibration \cite{Klemm,AspinwallTASI}.  Once again, this ubiquity of $K3$ is
presumably a consequence of the interpretation of the heterotic string as
the $K3$ wrapping of a fivebrane. Consequently, if $X$ admits {\it two}
different $K3$ fibrations, this would provide an alternative explanation
for heterotic dual pairs in four dimensions
\cite{Gross,Morrison1,Morrison2,AspinwallTASI} and this is indeed the case
for the Calabi-Yau manifolds discussed above.

\section*{Acknowledgements}

I have enjoyed valuable conversations with my colleagues at Texas A\&M:
Hong Lu, Jian Xin Lu, Sudipta Mukherji, Bengt Nilsson, Chris Pope, Ergin
Sezgin and Per Sundell.


\end{document}